\documentclass[prc,twocolumn,letterpaper,10pt,twoside,tightenlines,nofootinbib,showpacs,notitlepage,superscriptaddress
]{revtex4-1}
\usepackage{amsmath,amsfonts,amssymb,latexsym}
\usepackage{graphicx}
\usepackage[sort&compress]{natbib}
\usepackage{epstopdf}
\usepackage{subfigure}
\graphicspath{{./plots/}}

\newcommand{\m}{\cdot}

\newcommand{\n}{\nonumber \\}
\newcommand{\nn}{\nonumber}

\newcommand{\llkl}{\left\langle}
\newcommand{\rrkl}{\right\rangle}

\newcommand{\kl}{\left(}
\newcommand{\kr}{\right)}

\newcommand{\dd}{\mathrm{d}}

\newcommand{\lv}{\left\vert}
\newcommand{\rv}{\right\vert}

\newcommand{\del}{\partial}

\usepackage{slashed}
\usepackage{tikz}

\makeindex

\begin{document}

\title{Nonequilibrium photon production in partonic transport simulations}
\author{Moritz Greif}
\email{greif@th.physik.uni-frankfurt.de}
\affiliation{Institut f\"ur Theoretische Physik, Johann Wolfgang Goethe-Universit\"at,
Max-von-Laue-Str.\ 1, D-60438 Frankfurt am Main, Germany}
\author{Florian Senzel}
\author{Heiner Kremer}
\author{Kai Zhou}
\author{Carsten Greiner}
\affiliation{Institut f\"ur Theoretische Physik, Johann Wolfgang Goethe-Universit\"at,
Max-von-Laue-Str.\ 1, D-60438 Frankfurt am Main, Germany}
\author{Zhe Xu}
\affiliation{Department of Physics, Tsinghua University, Beijing 100084, China}
\affiliation{Collaborative Innovation Center of Quantum Matter, Beijing, China}
\date{\today }

\begin{abstract}
We discuss the implementation of leading-order photon production in  nonequilibrium partonic transport simulations. In this framework photons are produced by microscopic scatterings, where we include the exact matrix elements of Compton scattering, quark-antiquark annihilation, and bremsstrahlung processes. We show how the hard-thermal loop inspired screening of propagators has to be modified such that the microscopic production rate agrees well with the analytically known resummed leading-order rate.
We model the complete quark-gluon plasma phase of heavy-ion collisions by using the partonic transport approach called the Boltzmann approach to multiparton scatterings (BAMPS), which solves the ultrarelativistic Boltzmann equation with Monte Carlo methods. We show photon spectra and elliptic flow of photons from BAMPS and discuss nonequilibrium effects. Due to the slow  quark chemical equilibration in BAMPS, the yield is lower than the results from other groups; in turn we see a strong effect from scatterings of energetic jet-like partons with the medium. This nonequilibrium photon production can dominate the thermal emission, such that the spectra are harder and the photonic elliptic flow 
of the quark-gluon plasma becomes negative.
\end{abstract}

\maketitle

%%%%%%%%%%%%%%%%%%%%%%%%%%%%%%%%%%%%%%%%%%%%%%%%%%%%%%%%%%%%%%
%%%%%%%%%%%%%%%%%%%%%%%%%%%%%%%%%%%%%%%%%%%%%%%%%%%%%%%%%%%%%%
% A B S T R A C T
%%%%%%%%%%%%%%%%%%%%%%%%%%%%%%%%%%%%%%%%%%%%%%%%%%%%%%%%%%%%%%
%%%%%%%%%%%%%%%%%%%%%%%%%%%%%%%%%%%%%%%%%%%%%%%%%%%%%%%%%%%%%%

%%%%%%%%%%%%%%%%%%%%%%%%%%%%%%%%%%%%%%%%%%%%%%%%%%%%%%%%%%%%%%
%%%%%%%%%%%%%%%%%%%%%%%%%%%%%%%%%%%%%%%%%%%%%%%%%%%%%%%%%%%%%%
% I N T R O D U C T I O N
%%%%%%%%%%%%%%%%%%%%%%%%%%%%%%%%%%%%%%%%%%%%%%%%%%%%%%%%%%%%%%
%%%%%%%%%%%%%%%%%%%%%%%%%%%%%%%%%%%%%%%%%%%%%%%%%%%%%%%%%%%%%%

\section{Introduction}
\label{sec:Intro}
Photons have been used for decades as a valuable probe of the hot matter created in heavy-ion collisions.
Such matter, as created, e.g., in Au + Au collisions at the Relativistic Heavy Ion Collider (RHIC) at BNL or in Pb + Pb collisions at the Large Hadron Collider (LHC) at CERN, is highly dynamic, and temporarily the energy density is high enough that a so-called quark-gluon plasma (QGP) is formed \cite{Arsene:2004fa,Adcox:2004mh,Back:2004je,Adams:2005dq}. Photons are emitted from the initial nucleon-nucleon contacts (prompt photons), during the subsequent QGP phase and the hot hadron gas (HG) phase (thermal photons and jet-medium photons), by the fragmentation of jets outside the fireball, and finally  by the decay of long-lived resonances into real photons. The sum of all but the latter sources is called the \textit{direct} photon contribution, and experiments have succeeded in separating decay from direct photons (ALICE experiment at the LHC \cite{Adam:2015lda}, PHENIX experiment at RHIC \cite{Adare:2008ab,Afanasiev:2012dg,Adare:2014fwh_RHIC}). The measurements extend down to transverse momenta  $p_T=0.4\ (0.9)~\mathrm{GeV}$ for RHIC (LHC), and both find an exponential excess above $N_\text{coll}$-scaled prompt photons, which indicates a strong additional source, most likely the shining QGP and hot HG. The decay background subtraction is done via different methods, and improvements of the direct photon data are  expected in the future.\newline
Recently, ALICE and PHENIX have measured elliptic and triangular flow of direct photons for several centrality classes (PHENIX, $\sqrt{s}=200~\mathrm{GeV}$: $0\%-20\%,20\%-40\%,40\%-60\%$ \cite{Adare:2015lcd_RHICv2}, ALICE,$\sqrt{s}=2.76~\mathrm{TeV}$: $0\%-40\%,$ \cite{Lohner:2012ct_ALICEv2}). Both experiments show unexpectedly large flow; however, the measurement is extremely challenging and error bars are still large.\newline
It is nearly impossible for  experiments to disentangle the measured time-integrated photon spectra into their separate sources. Theoretical models however, compared with data, do not suffer from this problem. The ultimate goal is the explanation of the measured photon spectra by the correct combination of photon production mechanisms of hard and soft quantum chromo or electro dynamical (QCD or QED) interactions and a suitable spacetime evolution of the high-energy heavy-ion collision.\newline
It is furthermore desirable to explain the elliptic and triangular flow of photons in theoretical models. The explanation of elliptic flow for hadrons has required accurate modeling of the initial state and a correct treatment of the nearly hydrodynamic expansion of the medium with suitable viscosity \cite{Gale:2012rq,Schenke:2010rr,Qiu:2011iv,Werner:2010aa,Holopainen:2010gz}. It is crucial to also describe the flow of photons; however, its physical picture is substantially different. Photons leave the fireball without any further scattering such that their flow originates solely from the production process. For now, the large elliptic flow of photons poses a formidable challenge for dynamical models, and the simultaneous description of the yield and the flow of direct photons remains an unsolved puzzle. 

Until now, popular descriptions of the spacetime evolution of heavy-ion collisions are given by 
fireball parametrizations \cite{vanHees:2011vb, Rapp:2013nxa} or hydrodynamic simulations \cite{GALE2013,Schenke2011,PhysRevC.85.054902,Kolb:2003dz,Teaney:2001av,DelZanna:2013eua,Karpenko:2013wva,Holopainen:2010gz}. Photon spectra can be obtained from those models by folding the spacetime evolution of temperature $T$ and four-velocity $u^\mu$ over analytically known photon production rates $R(T,u^\mu)$ \cite{Paquet:2015lta,Chatterjee1long,Chatterjee2long,Liu:2008eh,vanHees:2011vb,Shen:2016zpp,Gale:2014dfa}.

Transport approaches, such as the Boltzmann approach to multiparton scatterings (BAMPS)~ \cite{Xu2005}, parton-hadron-string dynamics (PHSD) \cite{Cassing:2009vt,Linnyk:2013wma} or \textsc{urqmd} \cite{UrQMD1,UrQMD2} have two possibilities to study photon or dilepton production: ''coarse-graining" of the particle ensemble \cite{Endres:2015fna} and obtaining a spacetime background which can be used in the same way as a hydrodynamic evolution as described above, or by using the microscopic cross sections for the desired photon production processes and generating photons within the transport framework directly. The latter method will be our choice in the \textit{Boltzmann approach to multiparton scatterings} (BAMPS) \cite{Xu2005,Bouras2010a,Bouras2012,Bouras:2009nn,Fochler2010,Fochler2011,Uphoff2011a,Wesp2011,Reining2012,Uphoff:2012gb,Fochler2013,Greif2013,Senzel2013,Uphoff2013}, which is based on the numerical solution of the Boltzmann equation (BE).
 
We show how tree-level and radiative scattering diagrams can be implemented in dynamical transport simulations to nearly reproduce full leading-order (LO) photon rates. Subsequently we compute results for the QGP phase of high-energy nuclear reactions. The physical difference of our approach compared to hydro, fireball, or coarse-graining approaches is the intrinsic nonequilibrium nature - high or low energetic jets and the non, nearly, or full thermal  medium is treated equally. Furthermore, spacetime-dependent quark and gluon fugacities\footnote{For high-energy reactions the number of quarks and antiquarks is very similar, so that it makes sense to speak of an absolute quark fugacity defined as $\lambda_{q}\equiv n_{q+\bar{q}}/n_{q+\bar{q}}^{\text{equilibrium}}$ with the density $n$.} influence the photon rates by default. 

As a main result, we claim that the photon yield of the QGP can be much smaller than previously thought, due to the small initial quark content of the fireball. Furthermore, the pre-equilibrium phase of the QGP does not contribute significantly to yield or elliptic flow of direct photons. Second, we show how important nonequilibrium photon production can be for the elliptic flow: energetic particles behave ''jet-like``, and make the elliptic flow for higher transverse momenta negative. These results provide necessary complementary aspects to hydrodynamic calculations, which in most cases does not include strong off-equilibrium dynamics. 

Recently, much work is done concerning alternative rates (see, e.g.,~\cite{Iatrakis:2016ugz}) or rather ignored effects, such as viscous corrections (e.g.,~\cite{Vujanovic:2016anq}) or unknown sources (e.g.,~\cite{Benic:2016yqt}).\newline
This paper is organized as follows: In Sec.~\ref{sec:BAMPS} we describe our theoretical and numerical transport framework. In Sec.~\ref{sec:PhotonProduction} we introduce our implementation of $2 \leftrightarrow 2$ and $2 \rightarrow 3$ (radiative) photon production processes, their comparisons with hard thermal loop (HTL) resummed rates and explain the handling of interference effects. We discuss the scaling behavior with  quark fugacities in Sec.~\ref{sec:NontrivialFugacities}. In Sec.~\ref{sec:BoxTestJets} we show qualitatively in a box model how jet-medium interactions and a flowing thermal medium compete for elliptic flow and clarify the term jet-photon conversion in Sec.~\ref{sec:JetPhotonConversion}. Section~\ref{sec:results} is devoted to results for transverse momentum spectra and elliptic flow of photons from the QGP phase and its physical implications. Finally, we conclude in Sec.~\ref{sec:conclusion} and give an outlook on possible next steps.
Our units are $\hbar =c=k=1$; the spacetime metric is given by $g^{\mu \nu
}=\text{diag}(1,-1,-1,-1)$. Greek indices run from $0$ to $3$.

\section{The partonic cascade Boltzmann Approach to Multiparton Scatterings}
\label{sec:BAMPS}
We simulate the partonic evolution of heavy-ion collisions by using the $(3+1)$-dimensional transport approach (BAMPS) which solves the relativistic Boltzmann equation by Monte Carlo techniques \cite{Xu2005,Xu:2007aa} for  on-shell quarks and gluons by using perturbative QCD (pQCD) scattering matrix elements including $2\leftrightarrow 2$ and $2\leftrightarrow 3$ (radiative) processes.
With the phase-space distribution function $f^i(x,k)\equiv f_\textbf{k}^i$ for particle species $i$, the BE reads
\begin{align}
k^\mu \frac{\del}{\del x^\mu}f^i_\textbf{k} =\mathcal{C}^{2\rightarrow2}[f]+\mathcal{C}^{2\leftrightarrow3}[f],
\end{align}
 where $\mathcal{C}^{2\rightarrow2}[f]$ and $\mathcal{C}^{2\leftrightarrow3}[f]$ are the elastic and inelastic collision terms. BAMPS uses the test particle method: The physical particle number is increased by an integer factor $N_\text{test}$; however, all cross sections $\sigma$ are simultaneously scaled down, $\sigma\rightarrow\sigma/N_\text{test}$. This procedure increases the statistics but does not affect the physical results. Throughout this work, we include three flavors of light quarks, antiquarks, and gluons. All particles are on shell and massless (corresponding to an ideal equation of state) and carry physical electric charges and degeneracies. We neglect heavy quarks (see Refs.~\cite{Uphoff:2012gb,Uphoff:2011ad,Uphoff:2010sh}) because their presence is subdominant for photon observables.
Space is discretized in small cells with volume $\Delta V$ and particles scatter and propagate within time steps $\Delta t$. Within each cell, the probability for binary scattering is
\begin{equation}
P_{22}=\frac{\sigma_{\text{tot},22}(s)}{N_{\text{test}}}v_{\text{rel}}\frac{\Delta t}{\Delta V},
\label{eq:collProb}
\end{equation}
where $\sigma_{\text{tot},22}(s)$ is the (in general Mandelstam-$s$-dependent) binary total cross section. For $2\rightarrow 3$ particle scattering the probability is equivalently
\begin{equation}
P_{23}=\frac{\sigma_{\text{tot},23}(s)}{N_{\text{test}}}v_{\text{rel}}\frac{\Delta t}{\Delta V}.
\end{equation} 
The inelastic $3\rightarrow 2$ backreaction has a similar probability expression\footnote{We do not include $3\rightarrow 2$ processes involving photons, because these are subdominant processes. For gluon radiation it is implemented.}. For massless particles, the relative velocity of the two incoming particle with four-momenta $p_{1,2}=(E_{1,2},\vec{p}_{1,2})$ is $v_{\text{rel}}=s/(2E_1E_2)$.
For binary collisions, the cross sections are obtained via tree-level pQCD matrix elements, where propagators are ``screened'' by a LO HTL Debye mass (for photon production, see Sec.~\ref{sec:PhotonProduction}). For gluon radiation in $2\rightarrow 3$ inelastic collisions we use the Gunion-Bertsch approximation for the matrix elements \cite{Gunion1982}, which was further improved in Ref.~\cite{Fochler:2013epa}, whereas for radiated photons we use the full QCD+QED matrix element. BAMPS features a running coupling $\alpha_s(Q^2)$, which is evaluated at the momentum transfer $Q^2$ of the respective scattering process \cite{Uphoff:2012gb}. With this setup, the nuclear modification factor and elliptic flow in heavy-ion collisions could simultaneously be described in a former study \cite{Uphoff:2014cba}. The framework allowed for several kinetic studies such as the determination of transport coefficients, heavy quarks, Mach cones, jet energy loss and momentum asymmetry, see e.g., Refs.~\cite{Wesp:2011yy,Greif2013,Greif2014,Xu:2007jv,Xu:2008av,Xu:2010cq,Xu:2007ns,Uphoff:2012gb,Fochler:2008ts,Fochler:2010wn,Bouras:2009nn,Senzel:2013dta,Uphoff:2014hza,Senzel:2016qau}.

\subsection{Radiative cross sections}
The bremsstrahlung process $q+q\rightarrow q+q+\gamma$ is an important ingredient to the LO photon rate (more details in Sec.~\ref{sec:PhotonProduction}), thus in the following we give details regarding the evaluation of the total cross section. This is similar to the method for gluon radiation done earlier.
Radiative processes (particles 1 + 2 $\rightarrow$  3 + 4 +  5) are described by the momentum labels $p_1,p_2,p_3,p_4$ and $p_5\equiv k$. All considered $2\rightarrow 3$ processes have an internal gluon propagator with momentum $q$. The rapidity in the center of momentum (CoM)  frame is defined as $y=1/2 \ln\left[ (E+p_z)/(E-p_z)\right]$, where $y$ is the rapidity of the radiated photon, and its energy is $\omega = k_\perp \cosh y$. The energy of the outgoing particle 3 is $E_3=q_\perp \cosh y_3$, with its rapidity being $y_3$. Here, $k_\perp, q_\perp$ are the momentum components perpendicular to the $z$ axis in momentum space. The angle between $\vec{q}_\perp$ and $\vec{k}_\perp$ is $\phi$.
The total cross section for radiative processes is defined as
\begin{align}
\sigma_{2\rightarrow 3}&=\frac{1}{2s}\int\frac{\dd^3 p_3}{(2\pi)^32E_3}\frac{\dd^3 p_4}{(2\pi)^3 2E_4}\frac{\dd^3 k }{(2\pi)^3 2E_k}\n
&\quad\times\ (2\pi)^4 \delta^{(4)}\kl p_1+p_2-(p_3+p_4+k) \kr \lv \mathcal{M}_{2\rightarrow 3}\rv^2\n
&=\frac{1}{256\pi^4}\frac{1}{\nu}\frac{1}{s}\int^{s/4}_0 \dd q_\perp^2 \int ^{s/4}_{k^2_{\perp,\text{min}}} \dd k_T^2 \int^{y_{\text{max}}}_{y_{\text{min}}} \dd y \int_0^\pi \dd \phi\n
&\quad\times\  \left\vert \mathcal{M}_{2\rightarrow 3} \right\vert^2 \mathcal{J}\left[s,q_\perp,k_\perp,\phi,y \right],
\label{eq:sigma23Tot}
\end{align}
with a symmetry factor $\nu=n!$ for $n$ identical final-state particles, the radiative matrix element $ \left\vert \mathcal{M}_{2\rightarrow 3} \right\vert^2$ and the Jacobian 
\begin{align}
\mathcal{J}\left[s,q_\perp,k_\perp,\phi,y \right]=\sum\left\lbrace \kl \frac{\del F}{\del y_3} \kr^{-1} \right\rbrace,
\end{align}
where the sum is over the roots of 
\begin{align}
F&=(p_1+p_2-p_3-k)^2\n
&=s-2\sqrt{s}\kl q_\perp \cosh y_3 + k_\perp \cosh y \kr + 2q_\perp k_\perp \cos \phi\n
&\ + 2q_\perp k_\perp \kl \cosh y_3 \cosh y - \sinh y_3 \sinh y \kr.
\end{align}
The lower integration limit $k^2_{\perp,\text{min}}>0$ is further explained in Sec.~\ref{sec:interferenceEffects}.
The limits in the rapidity of the outgoing photon $y_{\text{max}},y_{\text{min}}$ are functions of $k^2_{\perp,\text{min}}$, $k_\perp$ and $s$. For given coordinates $s, k_\perp, q_\perp, y, \phi$ we can unambiguously obtain four-momenta in the CoM frame $p_1,p_2,p_3,p_4,k$ to get the value of the matrix element at this point without any approximation. The bremsstrahlung matrix element will be discussed in Sec.~\ref{sec:radiativePhotonProduction}.
More details regarding these kinematics can be found in Ref.~\cite{Fochler:2013epa}.
%%%

\section{Photon production rate in partonic transport}
\label{sec:PhotonProduction}
The emission rate of photons from an equilibrated quark-gluon plasma of temperature $T$ at leading-order $\mathcal{O}\kl e^2g^2T^4 \kr$ was first determined in Refs.~\cite{Arnold:2001ba,Arnold:2001ms}. In nearly all phenomenological studies concerning photons in heavy-ion collisions, these rates are used and we will denote them as ``AMY'' rates. Because the full leading-order rate contains both  $2\leftrightarrow 2$ photon production (namely Compton-scattering and quark-antiquark annihilation) and higher-order processes, such as bremsstrahlung and inelastic pair annihilation including coherence effects, we implement the $2\leftrightarrow 2$ processes and the $2\rightarrow 3$ processes separately.
We emphasize, that the advantage of transport simulations lies in the use of microscopic rates. We do not rely on thermal distributions of incoming partons, any pair of partons can produce a photon (given that the process is kinematically and diagrammatically allowed). However, in this section we use thermal distributions to show the validity of the total photon production rates in the transport framework by comparing to analytically known thermal rates.
\subsection{$2\leftrightarrow 2$ processes for photon production}
The authors of Refs.~\cite{Kapusta:1991qp,Baier:1991em} have computed the $2\leftrightarrow 2$ contribution to the photon rate in the $k/T\gg 1$ limit. This limit could be dropped in Refs.~\cite{Arnold:2001ba,Arnold:2001ms}. Essentially, in the so-called HTL improved rate the momentum transfer $t$ in photon production matrix elements is split up into a soft region $t<t^\star$ and a hard region $t>t^\star$. The rate from the hard region is treated in a straightforward way by integrating the appropriate squared matrix elements $\lv \mathcal{M}_{\text{Compton}}\rv ^2$ and $\lv \mathcal{M}_{\text{Annihilation}}\rv ^2$ in the rate integral (see Appendix~ \ref{appsec:rates}). The soft region, $t<t^\star$, is treated differently, by using effective HTL vertices and propagators in the corresponding loop diagrams. In the end, both soft and hard contributions are added and turn out to be independent of $t^\star$. 
In principle, the $t<t^\star$ calculation of the dressed loop diagram corresponds to the kinetic ($t>t^\star$) computation of the rate while making the propagators effectively massive, using a mass of order $gT$. For the vertices, however, a similar interpretation would be difficult. 
Within the partonic transport model BAMPS, we deal only with vacuum matrix elements, essentially the same which are used in the $t>t^\star$ region of Refs.~\cite{Kapusta:1991qp,Baier:1991em}. The matrix element for Compton scattering $qg\rightarrow q\gamma$ reads
\begin{align}
\lv \mathcal{M}_{\text{Compton}}\rv ^2=\frac{16}{3}\pi^2\alpha\alpha_s \kl \frac{s^2+st}{s^2} + \frac{s^2+st}{u^2}\kr.
\label{eq:Com}
\end{align}
The matrix element for quark-antiquark annihilation $q\bar{q}\rightarrow g\gamma$ is
\begin{align}
\lv \mathcal{M}_{\text{Annihilation}}\rv ^2=\frac{128}{9}\pi^2\alpha\alpha_s \kl \frac{tu}{t^2} + \frac{tu}{u^2}\kr.
\label{eq:Ann}
\end{align}
where $s,t,u$ are the usual Mandelstam variables.
\subsubsection{Screening of soft momentum transfers}
\label{sec:screnningKappa}
To avoid a crude cut-off like $t^\star$ in the momentum integration within BAMPS, we dress the quark propagators with a thermal mass $m_{D,q}\sim gT$, motivated by the HTL effective propagators. This screening of infrared divergencies naturally has a large effect on the total photon rate (and also the differential one), and must be carefully investigated, which is the purpose of this section. 
In BAMPS we use the formulas from Appendix~\ref{appsec:Debye} to compute the Debye mass from the given (in general nonequilibrium) distribution functions. We  want to mention here the systematic uncertainty concerning the strong coupling entering the Debye mass. It can be fixed (e.g., $\alpha_s=0.3$) or a running $\alpha_s(Q^2)$, where the scale $Q^2$ has to be specified. In the commonly used electric scale $Q=a2\pi T$, the prefactor $a$ is not clear, but $\mathcal{O}(1)$. In former versions of BAMPS, $Q^2$ was taken to be the momentum transfer of the specific process, $Q^2=s,t,u$. Moreover, in Ref.~\cite{Peshier:2006ah} it was argued, that the coupling can  be evaluated at the Debye mass itself, 
$m_D^2=\frac{4\pi}{3}\alpha_s(m_D^2)\kl N_c+\frac{N_f}{2} \kr T^2$. To allow for comparison with other groups we set the coupling in this paper fix to $\alpha_s=0.3$ for the photon production, unless otherwise stated. Note, that the procedure from this section and the following Secs.~\ref{sec:corrDistributionFction} and \ref{sec:TuneKinel} is only strictly valid for fixed coupling. Recent hydrodynamical calculations of photon rates also keep the coupling fixed.
\begin{figure}
	\centering
	\subfigure[\  The parameter $\kappa=2.45$ is tuned to make both integrated rates equal. \label{pic:fixKappa}]{\includegraphics[width=0.95\columnwidth]{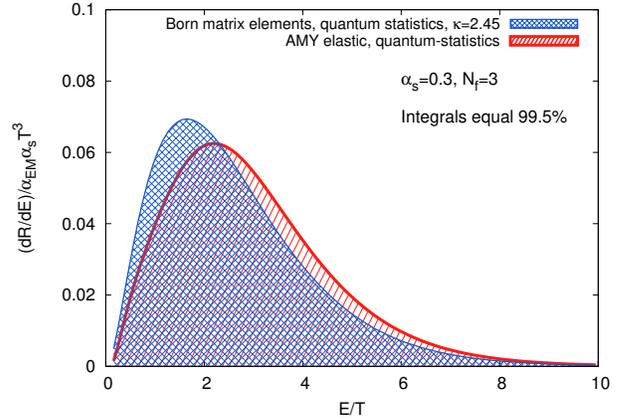}}\hfill
	\subfigure[\ The Born matrix element integrated with Boltzmann statistics (green dotted line). Reducing this rate by $C_{\text{stat}}=0.84$ (orange dashed line), the total rate $R$ equals the Born rate with quantum statistics, which equals approximately the elastic HTL improved  rate, see panel (a). \label{pic:fixCstat}]{\includegraphics[width=0.95\columnwidth]{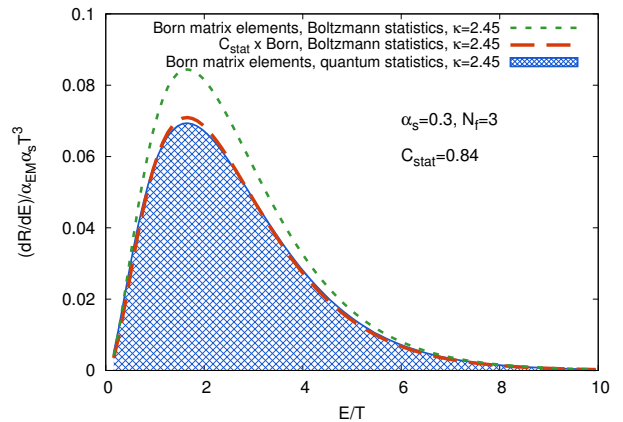}}
	\caption{The photon rate from Refs.~\protect\cite{Arnold:2001ba,Arnold:2001ms} compared  with the rate obtained from the numerical solution of Eq.~\eqref{eq:fullIntegralElastic} with matrix elements Eq.~\eqref{eq:Ann} and \eqref{eq:Com}, using a Debye mass $\kappa m_{D,q}^2$. In the top panel the $\kappa$ is fixed. In the bottom panel  we keep $\kappa=2.45$ and fix the parameter $C_{\text{stat}}$ by integrating the Born matrix element with Boltzmann statistics.}
\end{figure}
Having the screened matrix elements at hand, we then carry out the integration to obtain the total cross section and finally the photon spectra. These rates will by construction not be equal to the HTL improved rate, which is why we multiply the thermal masses by a real number $\kappa$. The propagators for the different channels read correspondingly,
\begin{align}
\frac{1}{t^2}&\rightarrow\frac{1}{(t-\kappa m_{D,q}^2)^2},\quad \frac{1}{u^2}\rightarrow\frac{1}{(u-\kappa m_{D,q}^2)^2},\n
\frac{1}{s^2}&\rightarrow \frac{1}{(s+\kappa m_{D,q}^2)^2}.
\label{eq:propagators}
\end{align}
\begin{table}
\centering
\begin{tabular}{|c|c|}
\hline 
Moment &  AMY/Born \\ 
\hline 
0'th & $99.5~\%$ \\ 
\hline 
1st & $112.5~\%$ \\ 
\hline 
2nd & $121.9~\%$ \\ 
\hline 
3rd & $128.1~\%$ \\ 
\hline 
4th & $132.1~\%$ \\ 
\hline 
\end{tabular} 
\caption{The comparison of AMY with Born-photon rates for higher moments of the photon rate, using the fixed value of $\kappa=2.45$.}
\label{tab:higherMoments}
\end{table}
It is now our strategy to choose the value of $\kappa$ in such a way that our simplified procedure leads to a rate that resembles the HTL improved rate closely (a similar procedure was done for heavy quark energy loss; e.g., in Ref.~\cite{Meistrenko:2012ju}). We do this by comparing the moments of the rate (where the $n$th moment is defined as $\int_0^\infty \dd E E^n \frac{\dd R}{\dd E }$). To this end we solve the integral in  Eq.~\eqref{eq:fullIntegralElastic} numerically (as in Appendix A of \cite{Shen:2014nfa}) first for quantum statistical distributions and screened matrix elements including the $\kappa$-factor (we call this ``Born'' rate), and compare the result to the HTL improved ($2 \leftrightarrow 2$) rate from \cite{Arnold:2001ba,Arnold:2001ms}. We adjust $\kappa$ [which is of order $\mathcal{O}(1)$] so that the total rates\footnote{The total rate is the total number of photons emitted per volume per time ($0$th moment), $R=\int_0^\infty \dd E \frac{\dd R}{\dd E}$.} $R$ are equal.
The comparison is shown in Fig.~\ref{pic:fixKappa}, where we plot $\dd R/\dd E$ in both schemes.
One observes that the Born rate (blue cross shaded area) has a slightly shifted peak when compared with the HTL improved rate. To get a handle on the quality of the comparison, we compare higher moments of the rate; the results are shown in Table~\ref{tab:higherMoments}. Note that the result for $\kappa$ is rather insensitive to the numerical integration limits, as the integrand drops to zero for $E/T\rightarrow 0,\infty$.
\subsubsection{Correction of the distribution functions}
\label{sec:corrDistributionFction}
\begin{figure}
\centering
\includegraphics[width=0.95\columnwidth]{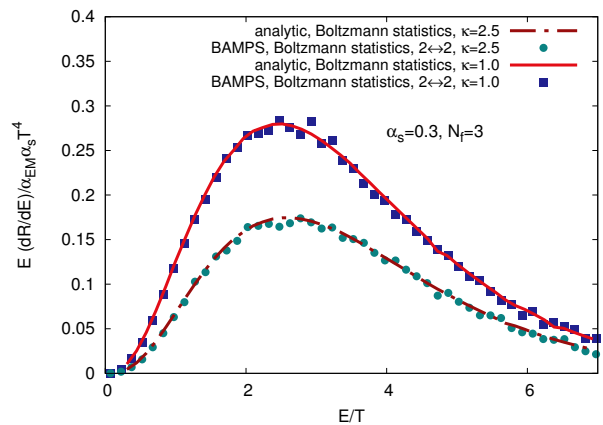} .
\caption{For two values of $\kappa$ we compare the numerically obtained $2\leftrightarrow 2$ photon rate to the analytic expectation (obtained by using the method from Ref.~\cite{Shen:2014nfa}).}
\label{BampsvsAnalytics}
\end{figure}

Finally, we need to correct for the small effect of the distribution functions. In the present numerical study, we can only use Boltzmann (classical) statistics, in initial and final states. There is no Pauli blocking or Bose enhancement \cite{Xu:2014ega}. That is why we will multiply the cross sections (equivalent to the rate) with a factor $C_{\text{stat}}$ in BAMPS to get the correct number of photons even without quantum statistics. This factor does not alter the differential cross section, as it is an overall prefactor. Note that also the Debye mass follows the Boltzmann distribution, because it is dynamically computed from the simulation.
To obtain $C_{\text{stat}}$, we solve Eq.~\eqref{eq:fullIntegralElastic} numerically with Boltzmann distributions and ignore Pauli blocking or Bose enhancement, but keep the fixed value of $\kappa$ from the procedure above\footnote{Here again, the Debye mass is in Boltzmann form.}. Then we compare again to the HTL resummed $2\leftrightarrow 2$ rate from Refs.~\cite{Arnold:2001ba,Arnold:2001ms}, which uses quantum statistics. The difference of both total rates is $C_{\text{stat}}$. 
The rates are shown in Fig.~\ref{pic:fixCstat}. The fact that $C_{\text{stat}}$ is below unity implies that the Pauli-blocking effect of the outgoing quark in the Compton channel is more important than the Bose enhancement effect of the outgoing gluon in the annihilation channel. This is consistent, because the Compton process happens more often (due to the combinatorics of the ingoing particles). 
Finally, we obtain the $2\leftrightarrow 2$ photon production rate from BAMPS  including the above explained ingredients in a box calculation. As an important numerical check, we compare the numerical results with the analytic expectation by using the exact same matrix elements (using two arbitrary values of $\kappa$ for illustration) in Fig.~\ref{BampsvsAnalytics}, and find excellent agreement.
\subsection{Radiative photon production}
\label{sec:radiativePhotonProduction}
\begin{figure}
  \centering
    \subfigure[$\ i\mathcal{M}_a,$\label{fig:matrixa}]{\includegraphics[width=0.2\textwidth]{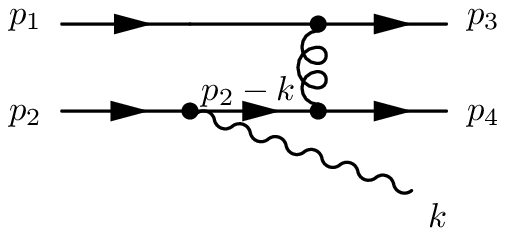}}\hfill
    \subfigure[$\ i\mathcal{M}_b.$\label{fig:matrixb}]{\includegraphics[width=0.2\textwidth]{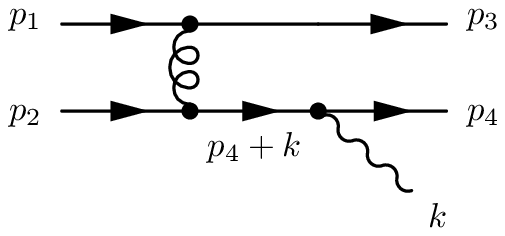}}\\
\caption{The two contributing vacuum diagrams we use for the numerical evaluation of the radiative photon rate. All internal propagators are screened by hand, and an overall factor $K_{\text{inel}}$ ensures the similarity to the AMY rate, as explained in Sec.~\ref{sec:TuneKinel}.\label{pic:twoloop_interference}}
\end{figure}
\begin{figure}
	\centering
	\subfigure[\label{bremsCutA}]{\includegraphics[width=0.25\textwidth]{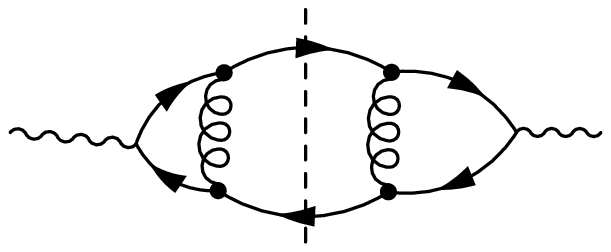}}\hfill
	\subfigure[\label{bremsCutB}]{\includegraphics[width=0.2\textwidth]{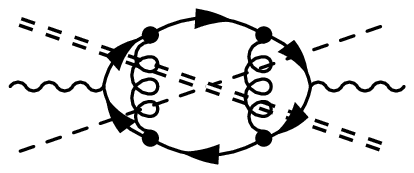}}\\
	\caption{An example of the photon self energy and its cuts to obtain scattering matrix elements along the method from Ref.~\protect\cite{Carrington:2002bv}\label{pic:photonSelfEnergy}. In diagram (a) the dashed line represents one possible cut, and the closed loops must be opened (''tic-ed'') to get a scattering matrix element. In (b) the dashed and double dashed lines are possible, topologically different cuts, generating Bremsstrahlung with on-shell gluons of the medium. \label{fig:photonSelfEnergy}}
\end{figure}

Motivated by the processes which give contributions to the total photon rate from Refs.~\cite{Arnold:2001ba,Arnold:2001ms}, we include radiative photon processes in BAMPS. We restrict ourselves to the simplest bremsstrahlung diagram,\footnote{Diagrams with more vertices become numerically very elaborate.} shown in Fig.~\ref{pic:twoloop_interference} with both subdiagrams. In Refs.~\cite{Arnold:2001ba,Arnold:2001ms} it is shown, that at leading-order in the rate, only the self-energy in the form of Fig.~\ref{fig:photonSelfEnergy} contributes, including a resummation of infinite gluon rungs. We want to stick to this picture, and neglect diagrams which would not emerge by cuts of this self-energy, even though in our transport setup those could be substantial. To this end, we employ the cutting-rules of Ref.~\cite{Carrington:2002bv}, and obtain scattering matrix elements. Every cut propagator is put on-shell, as well as every opened loop. In Fig.~\ref{bremsCutA} such a cut is shown, for the case of two gluon rungs. The loops must be opened by ``tics'' (see Ref.~\cite{Carrington:2002bv}) in every possible way. What emerges is exactly the diagrams of Fig.~\ref{pic:twoloop_interference}. Note that the cuts of Fig.~\ref{bremsCutB} produce two on-shell gluons, and one quark line radiating a photon. This, and corresponding diagrams with more gluon rungs, represent a sequential scattering with gluons, which is included by default in BAMPS, because the dominating subprocess $q+g\rightarrow q+g$ was included from the beginning, and the rare  radiation of the photon is Compton scattering in this case. 
Note that a (possible) $2\rightarrow 3$ process like $g+q\rightarrow q+g+\gamma$ (with a three gluon vertex), is not included in the set of diagrams resulting from the cuts. Ignoring the rigorous LO power counting of Refs.~\cite{Arnold:2001ba,Arnold:2001ms}, and just looking at the number of vacuum QCD vertices, this process could be included and would contribute significantly within BAMPS, because gluons are abundant, especially in the early phase of the QGP. This will be investigated in a future study. For now we use only one kind of matrix element, motivated by the leading-order picture.
As we only have vacuum matrix elements, we will insert thermal screening masses by hand into the propagators, as done before in the case of $2\leftrightarrow 2$ scattering. In Appendix~\ref{appsec:Bremsstrahlung} we derive the full squared matrix element $\overline{\lv\mathcal{M}\rv^2}_{\text{rad.}}$ starting from spinors and propagators without any further approximation. This is the radiative matrix element for photons that we will use in BAMPS, using techniques from Ref.~\cite{Fochler:2013epa}.
\subsubsection{Interference effects} 
\label{sec:interferenceEffects}
Photon radiation from bremsstrahlung processes suffers from the Landau-Pomeranchuk-Migdal (LPM) effect. The calculation of the radiative photon production rate in Refs.~\cite{Arnold:2001ba,Arnold:2001ms} fully includes the interferences among subsequently radiated photons. The notion of ``destructive interference'' of photons is motivated by looking at possible cuts of the retarded photon self energy and the resulting matrix elements. They must be summed and squared to obtain the full amplitude. In our microscopic description which is based on individual scatterings, we use an effective method to simulate the LPM interference effect. Within a transport approach, using individual scatterings for photon production, such interferences are necessarily destroyed, and must be restored by hand.

\begin{figure}[t!]
\centering
\includegraphics[width=0.6\columnwidth]{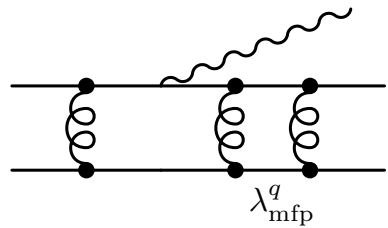}
\caption{Here we show a sketch of an LPM interference effect: Due to a short quark mean-free path a subsequent radiation is suppressed. Note that this diagram is not used as shown here, we rather evaluate the quark-quark elastic mean-free path dynamically in BAMPS and compare it to the formation time of the photon. The photon is produced by the pure bremsstrahlung subdiagram. \label{pic:LPM_MFP}}
\end{figure}

\begin{figure}
	\centering
	\includegraphics[width=0.95\columnwidth]{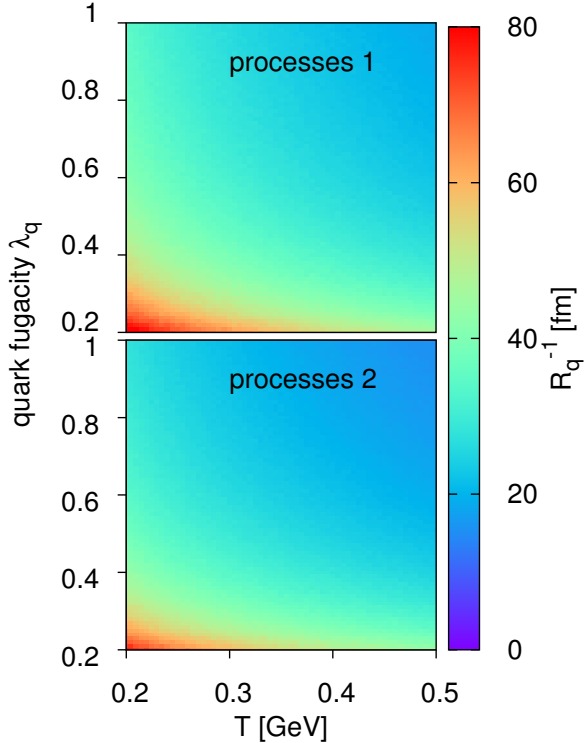}
	\caption{Numerical results for the specific mean-free path for a quark corresponding to different reactions, depending on temperature and fugacity.\label{fig:specMFP}}
\end{figure}

At first,  we calculate the specific inverse rate $\lambda^{\text{spec}}_{\text{mfp}}$ of the quark species which appear in the inelastic matrix elements for photon production [this can be seen as a mean-free path (mfp), where only certain scattering processes are included]. For the calculation of $\lambda^{\text{spec}}_{\text{mfp}}$ we take solely the specific $2\leftrightarrow 2$ processes into account which appear as subdiagram before or after the photon is radiated (see Fig.~\ref{pic:LPM_MFP}). These specific processes are:
\begin{description}
\item[processes 1] $qq\rightarrow qq$ / $\bar{q}\bar{q}\rightarrow \bar{q}\bar{q}$ 
\item[processes 2] $q\bar{q}\rightarrow q\bar{q}$
\end{description}
Here $q\ (\bar{q})$ are quark (antiquark) species, for up, down and strange quarks.
The corresponding numerical method is explained in Appendix~\ref{appsec:mfp}, and a typical process is schematically depicted in Fig.~\ref{pic:LPM_MFP}. 
In Fig.~\ref{fig:specMFP} we show numerical results for the inverse rate (mean-free path) corresponding to these processes separately. It depends strongly on the (anti-)quark fugacity and temperature.  
We will come back to the fugacity dependence of the mean-free path and the rate in Sec.~\ref{sec:NontrivialFugacities}.
 Next we multiply the amplitude for photon radiation by a Heaviside-function $\Theta\kl \lambda^{\text{spec}}_{\text{mfp}} - \tau_f \kr$ which ensures, that the formation time $\tau_f$ of the radiated photon is smaller than the mean-free path of the radiating quark,
\begin{align}
\overline{\lv\mathcal{M}\rv^2}_{\text{rad.}}\ \rightarrow\  \overline{\lv\mathcal{M}\rv^2}_{\text{rad.}}  \Theta\kl \lambda^{\text{spec}}_{\text{mfp}} - \tau_f \kr.
  \label{eq:rad.MatrixElement}
\end{align}
By doing this, we discard photons with such soft $k_\perp$ (transverse momentum relative to the radiating quark), that the radiating quark could have scattered again within the formation time. The $k^2_\perp$ integration in Eq.~\eqref{eq:sigma23Tot} in this case is limited by $k^2_{\perp,\text{min}} = \kl \lambda^{\text{spec}}_{\text{mfp}}\kr^{-2}$. As this procedure reflects the underlying interference effect only incomplete, we must insert a scale factor $K_\text{inel}$ in front of the matrix element.\newline
Recall that the current implementation of the LPM effect for radiated gluons in BAMPS is done in a similar way, the only difference is a factor $X_{\text{LPM}}$ being multiplied to the formation time and a different determination of the mean-free path. These differences are motivated by two physical effects: First, radiated gluons suffer from scattering after the radiation process, which dynamically alters their formation time. That is why we allow more radiated gluons than would actually be radiated if we required them to be fully formed. Second, gluon radiation rates involve far more diagrams (see, e.g., Ref.~\cite{Arnold:2002ja}), such that the mean-free path is the total mean-free path $qX\rightarrow Y$ where $X$ can be a quark or gluon.
\subsubsection{Fixing the scaling factor for bremsstrahlung}
\label{sec:TuneKinel}
As mentioned in the previous section, the implementation of radiative photon production is incomplete. There are in fact several parts which deviate from the AMY description. First, as we include only vacuum matrix elements with Debye screened propagators, we miss the correct treatment of soft momentum transfers. In the matrix element there are two propagators  (a quark- and a gluon propagator) where we insert Debye or thermal masses by hand and we could in principle tune these Debye masses by multiplying $\kappa$-factors as in the $2\leftrightarrow 2$ case from Sec.~\ref{sec:screnningKappa}. However, it is not clear how and if they should be tuned individually. The second simplification is the LPM effect described in the previous section. Third, we have the small effect of missing quantum statistics here, too. At last, the full AMY rate includes effectively not only the bremsstrahlung process, but also inelastic pair annihilation (a $3\rightarrow 2$ process), which we do not include here in this study. In Refs.~\cite{Arnold:2001ba,Arnold:2001ms} it is shown that this is a subdominant contribution. To cure all these problems, we scale the full matrix element $\overline{\lv\mathcal{M}\rv^2}_{\text{rad.}} $ with a factor $K_{\text{inel}}$. Such scaling is the simplest choice, and very common in transport approaches. 
\begin{figure}
\centering
\subfigure[\ The equilibrium radiative photon rate.\label{fig:AMYBAMPS}]{\includegraphics[width=0.95\columnwidth]{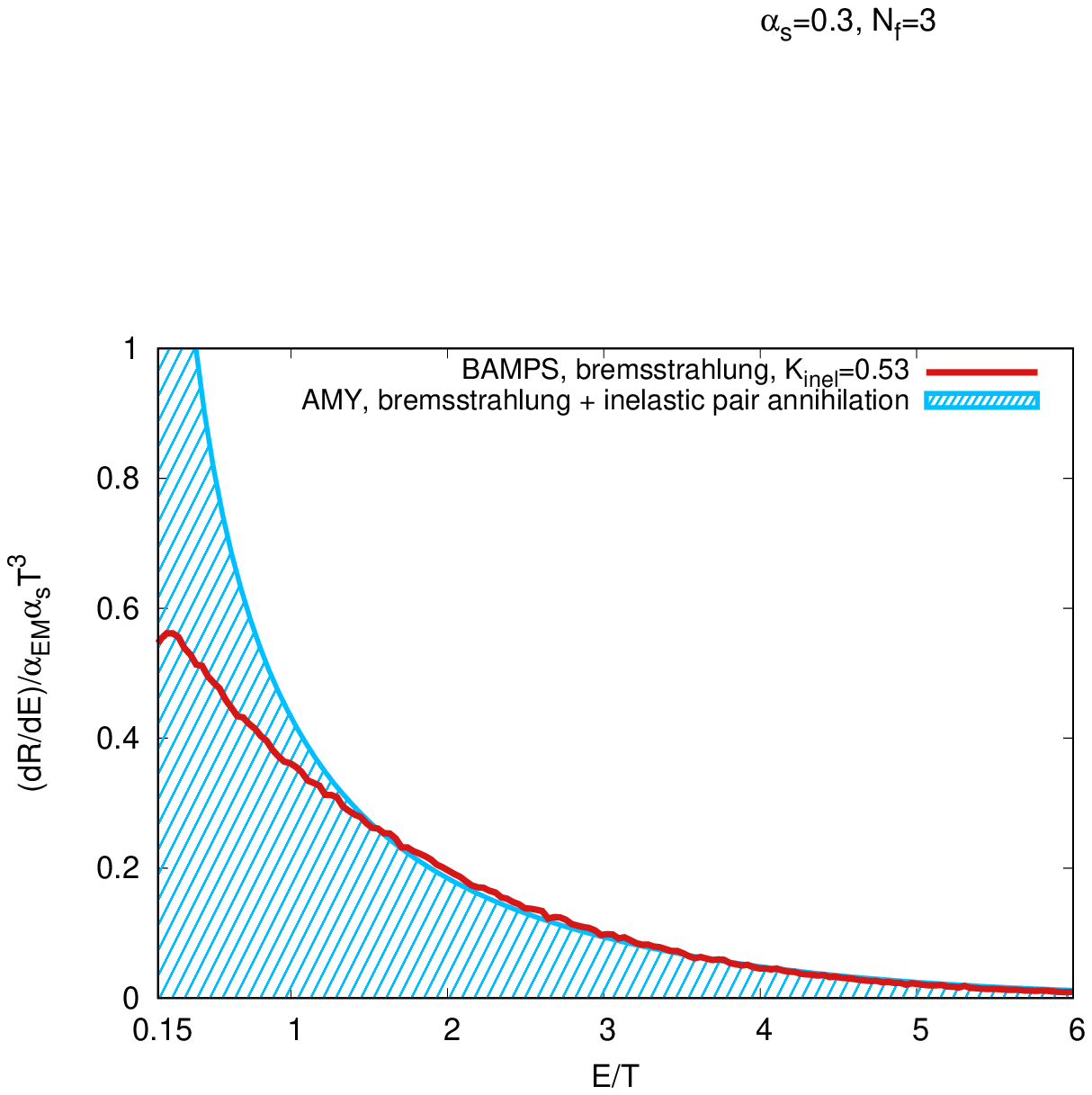}}
\subfigure[\ The equilibrium radiative photon rate weighted by photon energy.\label{fig:AMYBAMPS_firstmoment}]{\includegraphics[width=0.95\columnwidth]{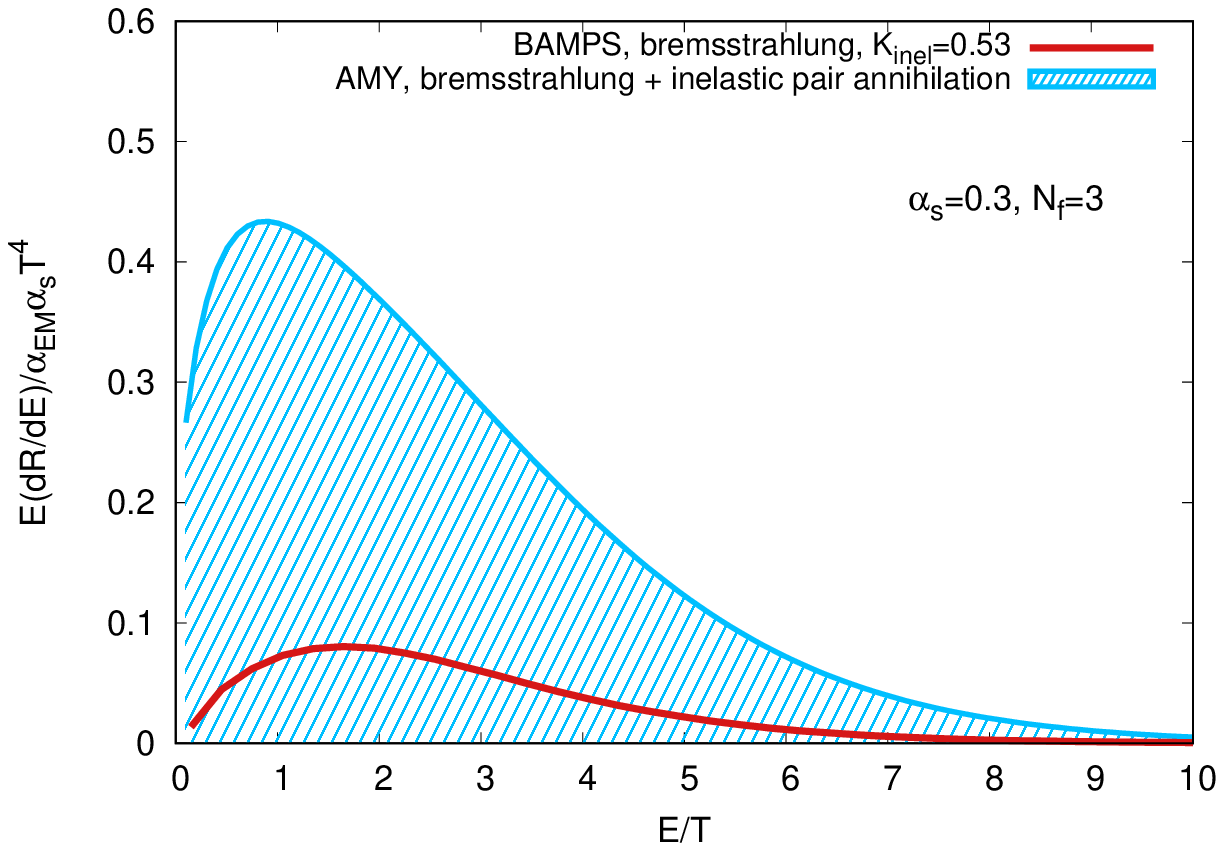}}\hfill\\
\caption{The photon rate for bremsstrahlung (a), and the rate weighted by the photon energy (b) compared to the full inelastic AMY result. The integral of the rate within $0.15<E/T<10$ is equal without a K-factor.\label{pic:AMY_BAMPS}}
\end{figure}
For inelastic processes, the total AMY rate $\mathrm{d}R/\mathrm{d}E$ diverges for small $E$, and the integral is ill defined. However, for small energies (transverse momentum), experiments do not measure anymore, and, the perturbative AMY description breaks down \cite{Ghiglieri:2016tvj}, so that we choose a suitable lower integration limit. To be consistent with the $2\leftrightarrow 2$ photon production we could choose to integrate over $0.15 < E/T < 10$, and compare the integral $R^{\text{inel}}$ to the result obtained by integrating the AMY result.  In this case we obtain $K_{\text{inel}}=0.79$. Having our application in mind, where we focus on transverse momenta in the range $0.5<p_T/\mathrm{GeV}<4$, we translate this at $T\sim 0.4~\mathrm{GeV}$ to a sensible integration region of $1<E/T<10$, where the result is very insensitive to the upper integration limit. For the following, we use this integration region, and obtain $K_{\text{inel}}=0.53$.
Using this factor, we make sure that we get (in an equilibrium case) the same number of photons and a similar spectrum in the energy region of interest.
In Fig.~\ref{pic:AMY_BAMPS} we show the numerical photon rate compared to the AMY rate, and also its first moment. The numerical rate from microscopic scatterings in Fig.~\ref{fig:AMYBAMPS} shows a similar slope as the AMY rate in the considered integration region, and the integrals of the curves in the plot are equal. The first moment in Fig.~\ref{fig:AMYBAMPS_firstmoment} from BAMPS is smaller than the AMY rate by about a factor of $5.2$, the second moment (not shown) by a factor of $5.3$. In Appendix~\ref{sec:verification} we show the corresponding differential cross sections and cross-checks of the kinematics. As a note, the thermal photon elliptic flow, being a transfer of flow from a boosted thermal distribution onto photons, is not sensitive to the differential photon rate (because photons are emitted isotropically in the local rest frame).

\subsection{Photon rate at nontrivial quark fugacities}
\label{sec:NontrivialFugacities}
The photon rate naturally depends on the quark and gluon content of the medium. For finite baryon chemical potentials (or quark chemical potential) the rate is modified by the (trivial) statistical factors ($q\bar{q}$ annihilation and Compton scattering behave differently), but also by other ingredients of the rate, such as the gluon self energies. These effects are studied thoroughly in Ref.~\cite{Gervais:2012wd}. The authors conclude, that the effect of the chemical potential to the photon spectra at RHIC or LHC is small, due to the small baryon chemical potential and the moderate sensitivity of the rates. Although we use a simplified diagrammatic setup, the effect of a quark-antiquark number asymmetry is included in the transport approach by default. For the present study at high energies, however, the effect is negligible.

\begin{figure}
\centering
\includegraphics[width=0.95\columnwidth]{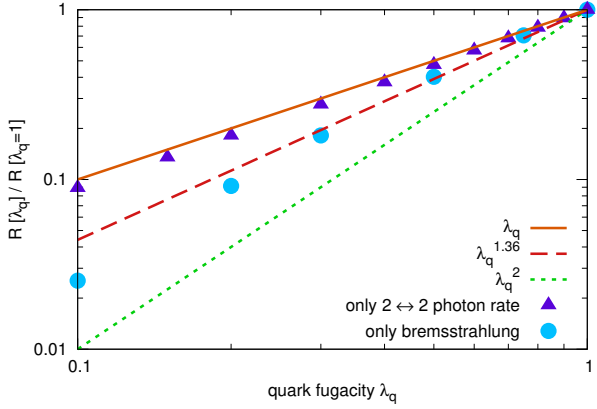}
\caption{Dependence of the photon rate $R$ on the  quark fugacity $\lambda_q$, and the  comparison to the naive expectations $R\sim \lambda_q^{1,2}$ (solid, dotted line). The bremsstrahlung rate shows roughly a $R\sim \lambda_q^{1.36}$ dependence (dashed line, fit), whereas the $2 \leftrightarrow 2$ photon production processes show a behavior $R\sim\lambda_q^{1.07}$.   \label{pic:FugScalingElasticInelastic}}
\end{figure}

\begin{figure}
\centering
\includegraphics[width=0.95\columnwidth]{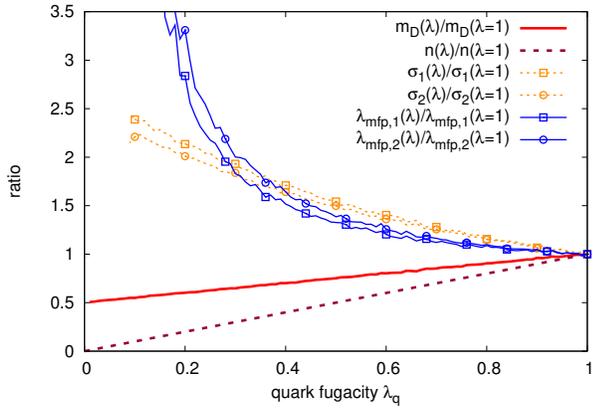}
\caption{Quark fugacity $\lambda_q$ scaling of Debye mass $m_D$, quark density $n$, average cross section at $T=0.4~\mathrm{GeV}$, $\sigma_i\equiv \llkl v_\mathrm{rel}\sigma_\mathrm{tot,i}(s) \rrkl$ (where $i=1,2,3$ corresponds to the three different specific processes considered) and the specific inverse rate $\lambda_{\mathrm{mfp}}$.   \label{pic:FugacScaling}}
\end{figure}

The second, more important characterization of the parton content is the ``absolute'' fugacity. Assuming by the previous argument, that the number of quarks equals the number of antiquarks, we define the gluon (quark) fugacities $\lambda_g(\lambda_q)$  as
\begin{align}
n_g &= \lambda_g n_g^{\text{equilibrium}}\n
n_q+n_{\bar{q}} &= \lambda_q \kl n_q^{\text{equilibrium}}+n_{\bar{q}}^{\text{equilibrium}} \kr. \nn
\end{align}
Effectively, for the considerations in this section, there is no difference between quark and antiquark. Note that the fugacities in heavy-ion collisions are in general time dependent.\newline
The initial state is still uncertain, especially the quark and gluon content is under debate. It is commonly believed, that gluons are saturated or over-saturated \cite{Stocker:2015nka}, and quark-antiquark pairs are not very abundant in the very early phase after the collision \cite{Xu:2007aa}. In Ref.~\cite{Srivastava:2016hwr} an undersaturation of quark-antiquark pairs ($\lambda_q<1$) seems to be favored by data within a rate equation approach. However, no precise answer about the fugacity dependence could be given up to now. Other studies \cite{Srivastava:1996rz,Stocker:2015nka,Vovchenko:2015yia,Shuryak:1992bt,Elliott:1999uz} give slightly different pictures, but we shall not elaborate on this topic here. Common ground is a quark fugacity $\lambda_q$ which is lower than unity and may or may not approach it within the lifetime of the fireball. We investigate in the following, how the photon rate behaves for nontrivial quark/gluon fugacities. Our arguments are similar to those of Ref.~\cite{Traxler:1995kx}.\newline
Naively, the $2\leftrightarrow 2$ Compton scattering (quark-antiquark annihilation) rates are proportional to $\lambda_q\lambda_g$ ($\lambda_q\lambda_q$) just by taking the incoming parton distribution functions into account. However, the Debye screening prescription from Eq.~\eqref{eq:propagators} lets the quark and gluon fugacities enter one more time into the rate. This will scale the rates differently as naively expected. In Fig.~\ref{pic:FugScalingElasticInelastic} we show the fugacity dependence of the $2 \leftrightarrow 2$ photon production (purple triangles) by comparing the total rate $R$ to the rate at unity fugacity, $R[\lambda_q]/R[\lambda_q=1]$. We have computed the Compton scattering and quark-antiquark annihilation rates for several quark fugacities (the gluon fugacity $\lambda_g$ is unity here), and find a combined scaling as $\lambda_q^{1.07}$. We conclude that the $2 \leftrightarrow 2$ rates can be seen as being simply proportional to the  quark fugacity. \newline
The inelastic photon rate will scale naively with $\lambda_q\lambda_q$; however, our implementation of the LPM effect uses the numerically (i.e., dynamically) evaluated quark mean-free path for specific processes (see Fig.~\ref{fig:specMFP}), which depends on the average cross sections $\sigma$ and particle densities $n$ and thus on the (quark) fugacity $\lambda_q$ as $\sim 1/(n\sigma)\sim 1/(\sigma\lambda_q T^3)$. The average cross sections are themselves Debye screened, and decrease for higher fugacities. These effects are summarized in Fig.~\ref{pic:FugacScaling}, where we show the scaling of Debye mass, density, average cross section and mean-free path, for the two processes  considered.\newline
Additionally, the fugacities enter also in the Debye screened gluon propagator. In Fig.~\ref{pic:FugScalingElasticInelastic} we show the scaling of the inelastic photon rate (normalized to the rate at $\lambda_q=1$) with the fugacity and compare with a naive scaling (without the effect from the LPM procedure or Debye screening), $R[\lambda_q]/R[\lambda_q=1]=\lambda_q^{2}$. By fitting a simple power law we find for bremsstrahlung roughly $R\sim \lambda_q^{1.36}$, for $\lambda_q\gtrsim 0.3$.

\section{Elliptic flow of photons originating from parton jets}
\subsection{Box calculation of photon leakeage effect}
\label{sec:BoxTestJets}
\begin{figure}
\centering
\includegraphics[width=0.95\columnwidth]{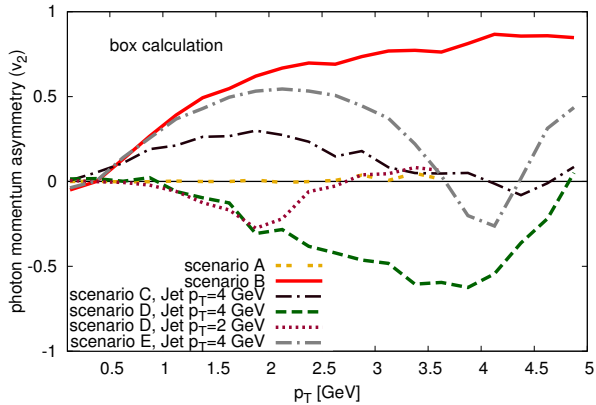}
\caption{\label{BoxTestJetsResults}Results for the qualitative understanding of elliptic flow of photons originating from flowing  thermal background and non thermal "jet"-like partons for the 5 scenarios explained in Sec.~\ref{sec:BoxTestJets}. The thermal medium has a temperature of $0.4~\mathrm{GeV}$, and for simplicity photons originate from $2 \leftrightarrow 2$ processes only.}
\end{figure}

To understand the kinetics of photons originating from hard partons qualitatively we use a fixed box with volume $V=L_x \m L_y \m L_z$, and populate it homogeneously with a thermal distribution of quarks and gluons (temperature $T$). This distribution can either be at rest with a four-velocity $u^\mu=(1,0,0,0)$, or boosted in the $x$ direction, $u^\mu=(\gamma,\gamma v_x,0,0)$, such that there is a strong collective flow in the $x$ direction (as seen from the laboratory frame). We change the box size to be either very thin, $L_x/L_y \ll 1$ or cubic, $L_x=L_y=L_z$. Furthermore, we initialize at the geometric center of the box a large amount of ``jet''-like particles isotropically with a fixed energy $E_j\sim 5T-10T$. All particles are allowed to scatter and produce photons, however, when any particle hits the wall, it is deleted. We define a transverse momentum, $p_T=\sqrt{p_x^2+p_y^2}$. Our observable resembles an elliptic flow $v_2$, but here it is merely a momentum anisotropy,
\begin{equation}
v_2=\llkl \frac{p_x^2-p_y^2}{p_T^2} \rrkl_{\text{average all photons}}.
\label{eq:v2}
\end{equation}
To this end, we consider five scenarios:
\begin{description}
\item[A] Cubic box at rest, including jets
\item[B] Cubic box with flow, without jets
\item[C] Cubic box with flow, including jets (jet $p_T=10 T$)
\item[D] Thin box, $L_x/L_y \ll 1$ at rest, including jets (jet $p_T=5T,10 T$)
\item[E] Thin box, $L_x/L_y\ll 1$ with flow, including jets (jet $p_T=10 T$)
\end{description}
Evaluating the momentum anisotropy from these scenarios, we plot the results in Fig.~\ref{BoxTestJetsResults}. As expected, no flow is visible in the symmetric scenario  A. In scenario B a thermal, flowing background generates a momentum anisotropy which increases for higher $p_T$. Undisturbed flow from the background is carried over to photons. Here we note that, by a simple relativistic effect, the (Lorentz variant) result of Eq.~\eqref{eq:v2} for produced particles is lower in magnitude than that for the background distribution. This effect depends on the boost. Including jets, which are isotropically emitted from the center, the flow reduces to zero at exactly the jet energy. For Compton scattering and quark-antiquark annihilation a large amount of photons inherit nearly the full momentum from the jets (jet-photon conversion). Because the jet momentum is dominant, the momentum anisotropy of these photons is zero, hence the curve of scenario C drops at the jet energy. The flow at lower $p_T$ stems from the background flow.
In scenario D there is no background flow, and no positive $v_2$ contribution. The jets, initialized in the middle of the box, traverse it until whichever wall comes first until they are deleted. During their traveling path, they can hit a thermal particle and produce a (conversion) photon, with a momentum close to that of the parent jet. This is more likely to happen in the (long) $y$ or $z$ direction, than in $x$, as the box has a small $L_x$ size. Most of the photons have larger $p_y$ momenta than $p_x$, thus the $v_2$ becomes negative (see, e.g., Ref~\cite{Turbide:2005bz} for similar findings). We show this effect for two different jet $p_T$ and, clearly, the minimal $v_2$ is reached at exactly the jet $p_T$. This effect can be termed the geometric leakage effect.  Finally, the combined effect of thermal background flow and jet conversion photons is shown in scenario E: For low $p_T$ there is substantial momentum anisotropy, whereas around the jet $p_T$ the conversion effect dominates and pushes the $v_2$ into the negative region.
This toy example shows what we can expect in a heavy-ion collision when both, jet particles and thermal flowing particles are present. The relative strengths of both effects have to be investigated in a full simulation.
\subsection{Jet photon conversion}
\label{sec:JetPhotonConversion}
\begin{figure}[t!]
\centering
\includegraphics[width=0.95\columnwidth]{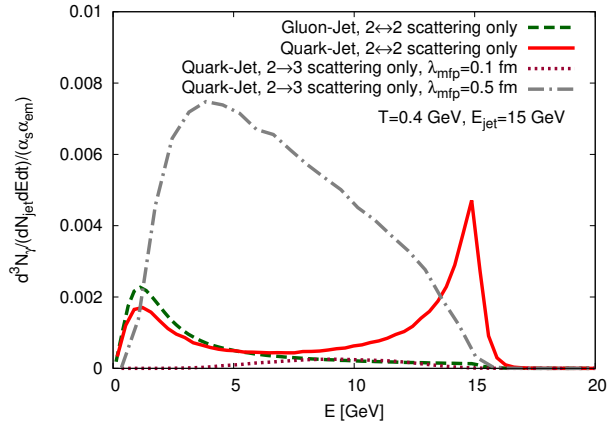}
\caption{\label{JetConversion} Spectra of photons which are produced by an incident jet particle with $E_{\text{jet}}=15~\mathrm{GeV}$ hitting a thermal bath. We show $2\leftrightarrow 2 $ and $2\rightarrow 3$ contributions separately.}
\end{figure}

To explicitly see how higher energetic partons (``jets'') interact with thermal particles and create a photon, we carry out a simple box calculation, where quark or gluon jets with fixed energy hit particles from a thermal bath. In Fig.~\ref{JetConversion} we show the resulting photon spectra, normalized by the volume density of jets $n_{\text{jet}}$. For a gluon jet, the only possible process is Compton scattering. It can be seen, that the photon spectrum is peaked at values $E\sim \mathcal{O}(T)$, due to the present channels. For gluons we cannot speak of jet-photon conversion. Quark jets, interacting only in $2\leftrightarrow 2$ processes (Compton scattering and quark-antiquark annihilation), have a dominating peak at the jet energy $E_{\text{jet}}=15~\mathrm{GeV}$. Due to momentum conservation the direction of the momentum of the photon must be very close to that of the jet quark - this is a true jet-photon conversion. The relative strength of the thermal peak at low energies and the peak at the jet energy depends on the ratio $T/E_{\text{jet}}$.
For bremsstrahlung we show the result for two different specific mean-free paths (dotted and dash-dotted line). The energy of the photon is distributed between the thermal scale and the jet energy scale, and depends on the LPM effect. However, in Appendix~\ref{sec:verification} we show that the differential cross section is peaked at low transverse momentum, which means the emission is favorably collinear to the jet quark. In this case we have a similar effect for the resulting photon as in the jet-photon-conversion case.  

\section{Results}
\label{sec:results}
In the following we show results from realistic simulations of heavy-ion collisions by using the photon production methods explained above within the framework of BAMPS. Details concerning the BAMPS setup for heavy-ion collisions can be found in \cite{Xu2005,Fochler:2008ts,Fochler:2010wn,Uphoff:2014cba}. The initial geometry of the collisions is governed by a  Glauber model \cite{Xu2005,Uphoff:2010sh}. For the initial parton distribution we use \textsc{pythia 6.4} \cite{Sjostrand:2006za}; details about the implementation can be found in Ref.~\cite{Uphoff:2010sh}. Because photons are very rare probes, they do not alter the collision dynamics. For this reason we use recorded BAMPS events, and sample photons by collisions among the recorded particles. This method allows us to enhance the photon cross section by a nearly arbitrary factor and scale the resulting spectra down by this factor (for better statistics). We have checked that all our results are independent of these factors. The background collision includes the latest improvements from BAMPS, such as the improved Gunion-Bertsch matrix elements for gluon radiation and a pQCD running coupling\footnote{Note that photon production is independent from the background events, and we chose the coupling to be fixed or running for the photon production cross sections, see also Fig.~\ref{SpectrumPT_RHIC_QGP_running}.}\cite{Uphoff:2014hza,Uphoff:2014cba,Fochler:2013epa}. The evolution of BAMPS runs until the energy density drops locally below $\epsilon_c=0.6~\mathrm{GeV}/\mathrm{fm}^3$. We have checked that the photon spectra are insensitive to this choice, because the rather cool medium in the later stages no longer produces many photons. 
\subsection{Photon yield from heavy-ion collisions} 
At present, BAMPS simulates only the QGP phase of heavy-ion collisions. This complicates studies and comparisons with photon data more than for other observables (such as, e.g., heavy quarks, jets, or bulk medium elliptic flow). 

%However, we attempt here to give some conclusive results, using in some cases additional results from other groups.\newline
\begin{figure}
\includegraphics[width=0.95\columnwidth]{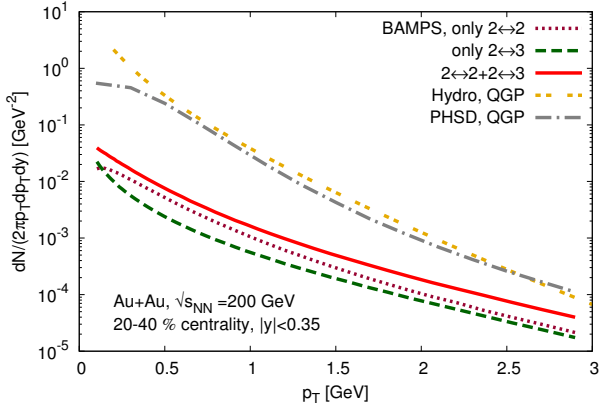}
\caption{\label{SpectrumPT_RHIC_QGP} The $p_T$ spectrum of direct photons from the QGP phase of Au + Au collisions at $\sqrt{s_{NN}}=200~\mathrm{GeV}$ for $20\%-40\%$ most central collisions. We show the elastic (magenta dotted) and inelastic (greed dashed) contribution from BAMPS as well as their sum (red solid) in comparison with a recent hydro result (yellow double dashed) from Ref.~\cite{Paquet:2015lta} and a result from off-shell transport PHSD \cite{Linnyk:2015tha}.}
\end{figure}

\begin{figure}
\includegraphics[width=0.95\columnwidth]{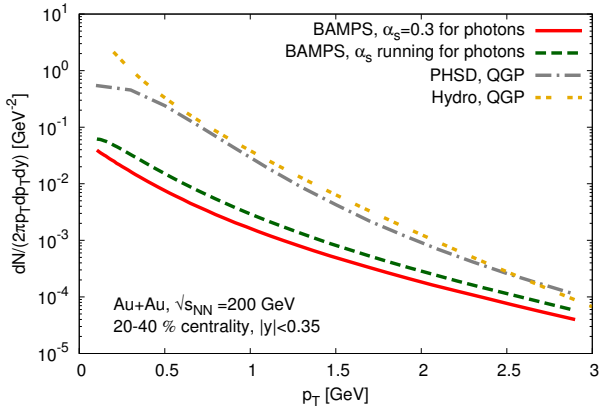}
\caption{\label{SpectrumPT_RHIC_QGP_running} Same as Fig.~\ref{SpectrumPT_RHIC_QGP}, but we switch on a running coupling for photon production (green dashed line).}
\end{figure}
\begin{figure}
\includegraphics[width=0.95\columnwidth]{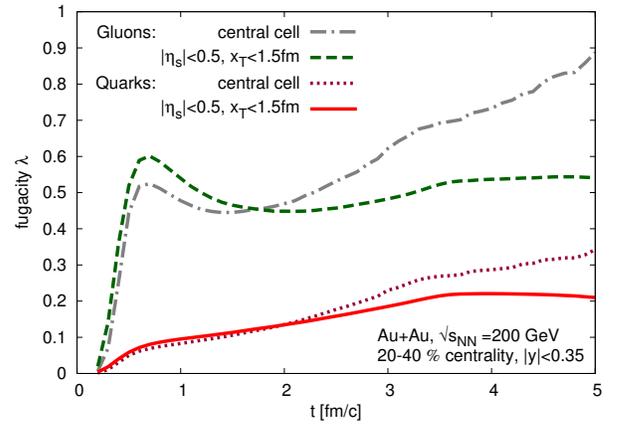}
\caption{\label{fugacitys} The average  quark and gluon fugacities over time in BAMPS for RHIC collisions at $\sqrt{s_{NN}}=200~\mathrm{GeV}$ at $20\%-40\%$ centrality. Shown is the average over only the central cell and a tube of transverse radius $1.5~\mathrm{fm}$ extending in spacetime rapidity $-0.5<\eta_s<0.5$.}
\end{figure}

In Fig.~\ref{SpectrumPT_RHIC_QGP} we show results for photon spectra in transverse momentum $p_T$ from BAMPS separately for $2\leftrightarrow 2$ photon production processes (magenta dotted line) and $2\rightarrow 3$ processes (green dashed line). The sum (red solid line) has an important contribution from the inelastic processes, especially at the highest and lowest $p_T$ . \newline
In Fig.~\ref{SpectrumPT_RHIC_QGP_running} we show the effect of a running coupling for photon production. The momentum transfer of the respective channel serves as scale $Q$ to evaluate the coupling, $\alpha_s(Q^2)$, but the coupling constant appearing inside the Debye masses is evaluated at the scale of an effective temperature in the corresponding cell ($Q=2\pi T_{\mathrm{eff}}$). The running coupling changes the slope only slightly, but increases the photon rate by a factor of $2$ below $p_T\lesssim 1.5~\mathrm{GeV}$ and $1.5$ above $p_T\gtrsim 1.5~\mathrm{GeV}$.\newline
Other models, such as PHSD \cite{Linnyk:2015tha} and MUSIC \cite{Paquet:2015lta} produce QGP rates around a factor of five to ten larger in magnitude than our results (for fixed $\alpha_s$), and a significantly steeper slope. The quark and gluon fugacities in MUSIC are unity, PHSD states only absolute particle numbers.
Due to the small yield of photons in the present setup, a possible
pre-equilibrium contribution from BAMPS to, e.g.,
hydrodynamic calculations is negligible. As the BAMPS results for
photons from the QGP undershoot
the hydrodynamic calculations for all $p_T$, and even hydrodynamics
underestimates experimental data, BAMPS
can not help in this direction with the present initial state. From all the above we see that the initial condition is the main uncertainty, and once more, our results underline the need to understand better the initial quark and gluon content of the fireball (see also Ref.~\cite{Oliva:2017pri}).
We show the fugacities in BAMPS in Fig.~\ref{fugacitys} for the same parameters of the collision. 
We have extracted the  quark fugacity by using an effective temperature for two representative geometries, the central cell of the collision and a tube of $1.5~\mathrm{fm}$ radius and length of one unit in rapidity. We remark that, at early times, these equilibrium quantities are only rough estimates of the quark content because the medium is not yet equilibrated.
As shown in Fig.~\ref{pic:FugScalingElasticInelastic} the $2 \leftrightarrow 2$ photon rates scale nearly linearly with the  quark fugacity, so that they are strongly affected by the  quark fugacities $\lambda_q\lesssim 0.2$ at early times in BAMPS. The inelastic rate has a more complicated fugacity dependence, such that the photon rate at $\lambda_q=0.2$ is less than $10\%$ of the equilibrium rate at $\lambda_q=1$. The combined effects explain the difference with the other models.

\begin{figure}
\includegraphics[width=0.95\columnwidth]{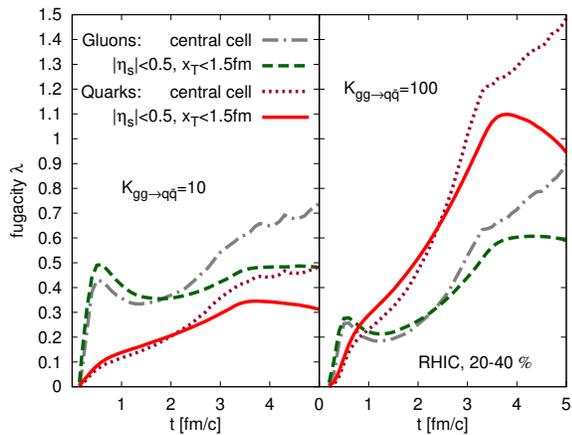}
\caption{\label{fugacitysK} Same as Fig.~\ref{fugacitys}, but for artificially increased quark-antiquark production cross section $\sigma_{gg\rightarrow q\bar{q}}=K \sigma_{gg\rightarrow q\bar{q}}$, where $K=10$ (left panel) and $K=100$ (right panel). The photon spectra are mostly sensitive to the early phase, where the notion of fugacity (or temperature) can only be effective.}
\end{figure}

\begin{figure}
\includegraphics[width=0.95\columnwidth]{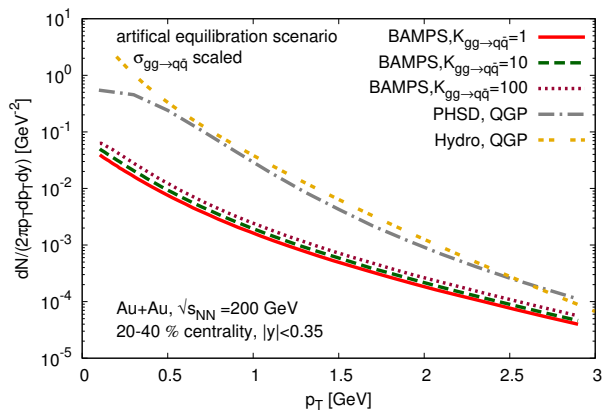}
\caption{\label{SpectrumPT_RHIC_QGP_K} Same as Fig.~\ref{SpectrumPT_RHIC_QGP}, but here we change the chemical equilibration during the evolution by artificially increasing the $gg\rightarrow q\bar{q}$ cross section by a factor of $10$ (magenta dotted line) and $100$ (green dashed line).}
\end{figure}

To see which role is played by the chemically equilibrating medium, we alter the fugacity evolution of the quarks (and thus also the gluons) by tuning arbitrarily the quark-antiquark production cross section\footnote{We ignore the tuning of the backreaction $q\bar{q}\rightarrow gg$ because the purpose of this test is to drive the chemical equilibration faster. In the central cell, the quark fugacity even increases above unity for late times and $K_{gg\rightarrow q\bar{q}}=100$.} by a factor of $10$ and $100$. The resulting fugacity evolution is shown in Fig.~\ref{fugacitysK}. It can be seen, that at around $t=2~\mathrm{fm/c}$ the quark fugacity increases from $\lambda_q(t=2~\mathrm{fm/c})\approx 0.15$ to $\lambda_q(t=2~\mathrm{fm/c})\approx 0.2$ (for $K_{gg\rightarrow q\bar{q}}=10$)  and $\lambda_q(t=2~\mathrm{fm/c})\approx 0.5$ (for $K_{gg\rightarrow q\bar{q}}=100$). 
In Fig.~\ref{SpectrumPT_RHIC_QGP_K} the resulting photon spectra are shown. The difference between the three scenarios is moderate, because most of the photons are produced within the first $2~\mathrm{fm/c}$. The difference in the fugacity is however, much stronger at later times (at $t=4~\mathrm{fm/c}$ about a factor of five), where not many photons are produced due to the thinner and colder medium. This shows, that the quark content at the very initial phase is crucial for photon spectra. Because the two other quoted models (MUSIC and PHSD) in Fig.~\ref{SpectrumPT_RHIC_QGP} and Fig.~\ref{SpectrumPT_RHIC_QGP_K} underestimate the data for RHIC slightly, our results suggest that this problem could be  even more severe. 

\subsection{Elliptic flow}
\label{sec:EllipticFlow}
\begin{figure}
\includegraphics[width=0.95\columnwidth]{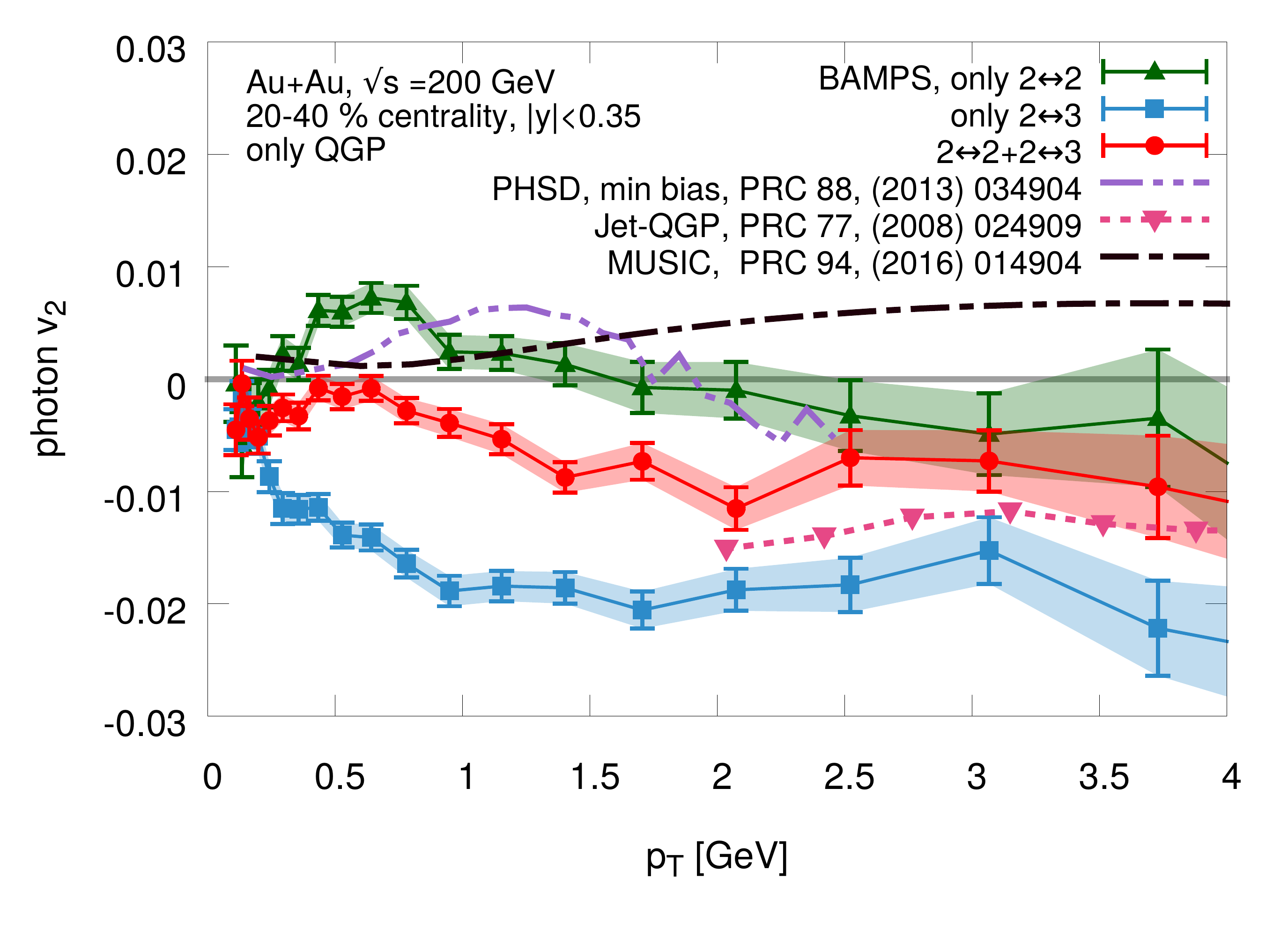}
\caption{\label{ellipticFlowBAMPS}The elliptic flow of photons from BAMPS for Au + Au collisions at $\sqrt{s_{NN}}=200~\mathrm{GeV}$ and $20\%-40\%$ centrality. Shown are the sum, and the elastic and inelastic contribution separately.}
\end{figure}

Within BAMPS, the event plane is known exactly, because we are dealing only with smooth Glauber initial conditions. This is the reason why elliptic flow can be conveniently obtained by averaging $(p_x^2+p_y^2)/p_T^2$ over all particles considered, photons in our case. Experimental results of direct photon elliptic flow are a weighted average over all sources of direct photons, weighted by their spectra. We can perform weighted averages by taking prompt photons and photons from hadronic scattering from elsewhere, in order to compare with data, but we find it instructive to compare directly the QGP contribution from BAMPS with other studies. In Fig.~\ref{ellipticFlowBAMPS} we show the elliptic flow of photons originating from only $2 \leftrightarrow 2$ collisions (green upward triangles), only bremsstrahlung (blue squares), and their sum (red points). 
The pink downward triangles~\cite{Turbide:2007mi} show elliptic flow of photons induced by jet-plasma interactions within a $(2+1)$-dimensional hydro model, where a time-dependent jet distribution is assumed and the jet-thermal rate is obtained by integrating separately the $2 \leftrightarrow 2$ and collinear (bremsstrahlung) rates using a thermal and a jet distribution. The final results are obtained by folding these rates over the hydro background. The thus-obtained elliptic flow is negative, a robust feature which was also observed in more simple one-dimensional (1D) Bjorken expansion \cite{Turbide:2005bz}, or with different initial conditions. Within BAMPS, we do not assume any jet distribution by hand, energetic particles propagate and suffer from energy loss by default. 
As we see in Fig.~\ref{ellipticFlowBAMPS}, at low $p_T$ the inelastic scattering shows only very little effect from the thermal flow, and its contribution is negative. The $2 \leftrightarrow 2$ photon production at $p_T\lesssim 1.3~\mathrm{GeV}$ shows a large flow, translated from the flowing background. The maximum magnitude is inline with the hydro result and the PHSD transport model.
The total photon flow in the QGP from BAMPS is very small in magnitude and negative.
We estimate the impact of our results when confronted to data by simply
replacing our QGP result with that from a complete hydro calculation and
compare roughly to experimental data.
The reweighting of the flow (the strongly flowing hadronic contribution
gets more weight, as our QGP yield is lower),
will (trivially) enhance the flow at lower $p_T$. At higher $p_T$ we see
the diminishing of flow due to the negative $v_2$ from BAMPS.
Again, bearing in mind the inconsistency, we have added a
``pre-equilibrium contribution'' from BAMPS to the complete hydro result.
This, as was the case also for the yield, has only a very small effect.

\section{Conclusions}
\label{sec:conclusion}
We have implemented photon production cross sections at full leading-order in a dynamical, microscopic transport approach for heavy-ion collisions. The conceptual difficulties concerning the rates, which involve in principle infinite scattering amplitudes, could be tackled by tuning the screening mass and fixing overall multiplicative factors. Consequently, the analytically fully known leading-order photon production rate has been reproduced by the transport simulation from microscopic scatterings. We discussed the Debye mass dependence as well as the fugacity dependence of the photon rate, and found a nontrivial scaling with fugacity, which is different for $2 \leftrightarrow 3$ and $2 \leftrightarrow 2$ photon production. Having the fugacity dependence of the photon rate under control, we turned to realistic heavy-ion collisions.

We give results for the direct photon contribution to spectra and elliptic flow from the QGP phase (in this exploratory study we restricted ourselves to RHIC collisions at $\sqrt{s_{NN}}=200~\mathrm{GeV}$). The magnitude of the $p_T$ spectrum naturally depends strongly on the quark content of the medium which in turn is largely influenced by the initial conditions. Our implementation of \textsc{pythia} initial conditions combined with the mentioned fugacity dependence has shown the expected smaller yield than hydrodynamic computations which is in complete chemical equilibrium from their initialization time on. The $p_T$ spectra from BAMPS are also harder; this is due to the choice of initial condition, but also a distinct feature of the nonequilibrium nature of BAMPS. Partons which are not part of a thermal ensemble scatter and make photons and these photons are not expected to show a thermal behavior. Especially at higher momenta (between $2-3~\mathrm{GeV}$), the spectrum is harder. If yet unknown initial conditions with larger quark content were used in the future, the nonequilibrium photon spectrum would be higher and thus closer to the data. In this case one would have a stronger pre-equilibrium contribution, whereas with our present setup the pre-equilibrium contribution is very small. 

A more obvious implication of nonequilibrium photon production can be seen in our results for the elliptic flow. Due to microscopic production of photons by partons which are not part of a pure thermal ensemble, the momentum asymmetry of the produced photons is not the result of a simple boosted thermal spectrum. Jet-photon conversion, the almost one-to-one transfer of momentum of (usually higher energetic) particles to photons, could have been identified to play an important role. We have observed the competing of a thermal flowing medium with positive photon $v_2$ and the nonequilibrium leakage effect with its negative $v_2$. This leakage effect probes the asymmetric geometry of the fireball by the traverse of slightly higher energetic quarks and their conversion into photons, or, radiative but very collinear emission. The resulting elliptic flow is dominated by nonthermal emission at higher $p_T$ and strongly negative, and larger but still negative for low $p_T$, where background flow is more important.
We believe that other yet unidentified effects solve the photon-puzzle in future. Fragmentation photons may play a role as well as the effect of electromagnetic fields on the evolution. Both effects can be investigated in microscopic transport models, and with this work we have set the basis.

\section*{Acknowledgements}
We are grateful to J.F. Paquet, C. Shen and C. Gale for constant interest in our work, useful discussions and providing data for comparison. Furthermore we thank Alexander Rothkopf for fruitful discussions.
M.G. and F.S. acknowledge the support from the ``Helmhotz
Graduate School for Heavy Ion research''. The authors are grateful to the Center for Scientific Computing (CSC) Frankfurt for the computing resources. This work was supported by the Helmholtz International Center for FAIR within the framework of the LOEWE program launched by the State of Hesse.
XZ was supported by the MOST, the NSFC under Grants No. 2014CB845400, No. 11275103, No. 11335005, and No. 11575092.

%%%%%%%%%%%%%%%%%%%%%%%%%%%%%%%%%%%%%%%%%%%%%%%%%%%%%%%%
% A P P E N D I X
%%%%%%%%%%%%%%%%%%%%%%%%%%%%%%%%%%%%%%%%%%%%%%%%%%%%%%%%
\appendix
 \section{Debye Screening Prescriptions}
 \label{appsec:Debye}
The screening masses and thermal quark masses behave very similarly for our purposes. They are of order $gT$ but have different prefactors depending on the type of statistics.
The squared thermal mass for light quarks is defined as
\begin{equation}
m_{D,q}^2=g^2C_F\int \frac{\dd^3 p}{(2\pi)^3E_p}(f_g+f_q).
\end{equation} 
The squared thermal gluon mass (=Debye mass) is defined as
\begin{align}
m_{D,g}^2=16\pi\alpha_s \int \frac{\dd^3 p}{(2\pi)^3 E_p}(N_c f_g+N_f f_q).
\end{align} 
Using Boltzmann statistic distributions, the squared gluon Debye mass is
 \begin{align}
 m_{D,g}^2=\frac{8}{\pi}(N_c+N_f)\alpha_s T^2\approx 15.28 \alpha_s T^2
 \end{align}
whereas the squared thermal quark mass is
\begin{align}
m_{D,q}^2=\frac{1}{9}m_D^2=\frac{8\alpha_s T^2}{9\pi}(N_c+N_f)=\frac{16}{3\pi}\alpha_sT^2\approx 1.7 \alpha_s T^2.
\end{align}
 Using quantum statistic distributions, the squared gluon Debye mass is
 \begin{align}
 m_{D,g}^2=\frac{4\pi\alpha_s}{3}\kl N_c + \frac{N_f}{2}\kr T^2 = 6\pi\alpha_sT^2 \approx 18.85 \alpha_s T^2,
 \end{align} 
whereas the squared thermal quark mass is
\begin{align}
m_{D,q}^2=\frac{1}{2}m_\infty^2=\frac{1}{2}\frac{C_Fg_s^2T^2}{4}=\frac{2\pi\alpha_s}{3}T^2\approx 2.09 \alpha_s T^2.
\end{align} 

\section{Photon Rates}
\label{appsec:rates}
The total photon production rate (units $[\mathrm{energy}^4]$) for processes $P+P^\prime\rightarrow K+K^\prime$, where $K$ is the four-momentum of the photon, can be written as \cite{Shen:2014nfa},
\begin{widetext}
\begin{align}
R&=\mathcal{N}\int \frac{\dd^3 p}{2E_p(2\pi)^3} \int \frac{\dd^3 p^\prime}{2E_{p^\prime}(2\pi)^3} \int \frac{\dd^3 k}{2E_k(2\pi)^3} \int \frac{\dd^3 k^\prime}{2E_{k^\prime}(2\pi)^3} (2\pi)^4\delta^{(4)}(P+P^\prime-K-K^\prime)\n
&\quad\times\lv \mathcal{M}\rv^2 f(P)f(P^\prime)\kl 1 \pm f(K^\prime) \kr,
\label{eq:fullIntegralElastic}
\end{align}
\end{widetext}
where $\mathcal{N}$ is a symmetry factor respecting the electric charges and degeneracies. In the case of Compton scattering, the symmetry factor for two flavors is $\mathcal{N}=320/3$, for three flavors $\mathcal{N}=128$. In the case of quark-antiquark annihilation, the symmetry factor for two flavors is $\mathcal{N}=20$, for three flavors $\mathcal{N}=24$.
By using an approximation for the case $E\gg T$, the differential photon rate can be obtained from the scattering matrix elements $\mathcal{M}(s,t)$ using \cite{Kapusta:1991qp}
\begin{widetext}
\begin{align}
E_k\frac{\mathrm{d}R_i}{\mathrm{d}^3k}=\frac{\mathcal{N}_i}{(2\pi)^6}\frac{T}{32E_k}e^{-E_k/T}\int\limits_0^\infty\mathrm{d}s \frac{1}{s} \ln\left\lbrace\kl 1\pm e^{-\frac{s}{4E_kT}} \kr^{\pm 1}\right\rbrace \int\limits_{-s}^{0} \mathrm{d}t \lv \mathcal{M}_i \rv^2,
\label{eq:hard_contr_w_approx}
\end{align}
\end{widetext}
where $s,\ t$ and $u=-s-t$ are the usual Mandelstam variables. However, by using the techniques from Ref.~\cite{Shen:2014nfa}, the rate can be integrated numerically without the approximation $E\gg T$.
Note that by using Eq.~\eqref{eq:fullIntegralElastic} or \eqref{eq:hard_contr_w_approx} the matrix element must not diverge for soft momentum transfer. These formulas can thus only be used, if either a soft momentum cutoff is applied ($q_{\text{cut}}$, as in most previous works, e.g., \cite{Shen:2014nfa}), or the propagators in the matrix elements are naively screened by using a screening mass. This we call Born approximation.
\section{Algorithm to Determine Specific Mean-free Paths}
\label{appsec:mfp}
The mean-free path is the inverse of the scattering rate per particle $\lambda_{\text{mfp}}=R^{-1}$.
The inverse rate for scattering of a single particle $q$ within a medium of particle density $n_q$ is
\begin{equation}
\lambda_{\text{mfp,}qq\rightarrow qq}^q=\kl n_q \llkl \sigma({s})v_{\text{rel}} \rrkl_\text{therm} \kr^{-1},
\label{eq:mfpFormula}
\end{equation}
where the average is over the thermal ensemble and $v_{\text{rel}}\equiv s/(2E_1E_2)$, where $E_1, E_2$ are the energies of two incoming particles and the Mandelstam variable $s=(P_1+P_2)^2$ is the squared sum over their four-momenta. A thermal ensemble allows for the direct calculation of the mean-free path from the thermal ensemble, just given the cross section $\sigma({s})$ and the equilibrium density $n_q$. However, we explicitly want to extract the mean-free paths in a chemical and/or kinetically nonequilibrated system. For this purpose, we choose all possible scattering partners $i$ in each computational cell and compute their collision probability $P^{i}_{22}$ from Eq.~\eqref{eq:collProb}, such that
 \begin{align}
\lambda_{\text{mfp,}qq\rightarrow qq}^q&=\kl n_q \llkl \sigma({s})v_{\text{rel}} \rrkl_\text{therm} \kr^{-1}\n
&=N_q\frac{1}{M} \sum\limits_{i=1}^M \frac{P_{22}^i}{\Delta t},\n
&=\frac{1}{\Delta V}\frac{2}{(N_q-1)}\sum\limits_{i=1}^M \frac{\sigma_i v_{\text{rel},i}}{\Delta t},\n
M&\equiv\frac{1}{2}N_q(N_q-1).
\label{eq:mfpFormulaNumerical}
\end{align}
Note, that here $N_q$ is the total number of quarks in the cell with volume $\Delta V$, which is the physical number times the number of test particles, $N_{\text{test}}$. The cross section in Eq.~\eqref{eq:collProb} is divided by $N_{\text{test}}$, such that the mean-free path is the physical mean-free path and independent of $N_{\text{test}}$.
For processes $qq\rightarrow qq$ there are ${N_q \choose 2}=1/2 N_q(N_q-1)$ possible scattering processes for $N_q$ quarks in the system, and we take numerically the average to get the mean-free path of a quark when considering only scatterings with another quark of the same flavor. In a similar way we can compute the mean-free paths for $qq^\prime\rightarrow qq^\prime,\ \bar{q}\bar{q}^\prime\rightarrow\bar{q}\bar{q}^\prime,\  q\bar{q}\rightarrow q\bar{q}$. 

\section{Bremsstrahlung Diagrams for Quark-Quark Scattering}
\label{appsec:Bremsstrahlung}
In this section we compute the squared matrix element for the $qq\rightarrow qq\gamma$ process, shown in Fig.~\ref{pic:twoloop_interference}. For this purpose, we label the amplitude of Fig.~\ref{fig:matrixa} with $\mathcal{M}_a$, and the one from Fig.~\ref{fig:matrixb} with $\mathcal{M}_b$. We have to compute $(\mathcal{M}_a+\mathcal{M}_b)^\star\m (\mathcal{M}_a+\mathcal{M}_b)$. As customary in scattering theory, the matrix element is given by an average over initial spin, polarization and color states, and a sum over final states. 
\subsection{Matrix elements}
With the momentum assignment $p_3=p_1+q$, $p_4=p_2-q-k$ we write down the first matrix element [Fig.~\ref{fig:matrixa}] by using momentum space Feynman rules:
\begin{align}
i\mathcal{M}_a&=\bar{u}^w(p_3)(ig)\gamma^\mu\lambda^a_{il}u^s(p_1)\frac{-ig_{\mu\nu}\delta_{ab}}{q^2}\bar{u}^r(p_4)\n
&\quad\times\ (ig)\gamma^\nu \lambda^b_{mj}\frac{i(m+\slashed{p}_2-\slashed{k})}{(p_2-k)^2-m^2}\n
&\quad\times\ (iQ_{EM})\gamma^\alpha \epsilon_\alpha^\star(k)u^t(p_2).
\end{align}
The second matrix element [Fig.~\ref{fig:matrixb}] is
\begin{align}
i\mathcal{M}_b&=\bar{u}^w(p_3)(ig)\gamma^\mu \lambda^a_{il} u^s(p_1) \frac{-ig_{\mu\nu}\delta_{ab}}{q^2}
\bar{u}^r(p_4)\n
&\quad\times\ (iQ_{EM})\gamma^\alpha\epsilon_\alpha^\star(k)\n
&\quad\times \frac{i(m+\slashed{p}_4+\slashed{k})}{(p_4+k)^2-m^2}(ig)\gamma^\nu\lambda^b_{mj}u^t(p_2).
\end{align}
By using the Dirac equation we transform the numerators of the quark propagators in the following way:
\begin{align}
(\slashed{p}_2+m)\gamma^\alpha u(p_2)=2p_2^\alpha u(p_2)\n
(\slashed{p_4}+m)\gamma^\nu u(p_2)=2p_4^\nu u(p_2),\nn
\end{align}
and we simplify the denominators,
\begin{align}
(p_2-k)^2=-2p_2\m k,\quad (p_4+k)^2=2p_4\m k.
\end{align}
Note that, later on, we screen the $t$-channel quark-propagator in $\mathcal{M}_a$ by using a Debye mass $m_{D,q}^2$,
\begin{align}
\frac{1}{-2p_2\m k}\rightarrow \frac{1}{-2p_2\m k -m_{D,q}^2},
\end{align}
and the $s$-channel propagator in in $\mathcal{M}_b$,
\begin{align}
\frac{1}{2p_4\m k}\rightarrow \frac{1}{2p_4\m k+m_{D,q}^2}.
\end{align}
Only at this step we set the masses to zero, $m\equiv 0 $. 
The gluon propagator will be screened with the Debye mass $m_{D,g}^2$,
\begin{align}
\frac{1}{q^2}\rightarrow \frac{1}{q^2-m_{D,g}^2}.
\end{align}
\subsection{Amplitude}
Next we simplify the summed matrix elements,
\begin{widetext}
\begin{align}
i\mathcal{M}_a+i\mathcal{M}_b&=\bar{u}^w(p_3)(ig)^2\gamma^\mu \lambda^a_{il}\lambda^b_{mj} u^s(p_1)\frac{-ig_{\mu\nu}}{q^2}(iQ_{EM})\bar{u}^r(p_4)\n
&\quad\times\left[  \frac{i(\gamma^\nu p_2^\alpha-\gamma^\nu\slashed{k}\gamma^\alpha)}{-2p_2\m k }   +    \frac{i(2\gamma^\alpha p_4^\nu+\gamma^\alpha\slashed{k}\gamma^\nu)}{2p_4\m k}     \right]u^t(p_2)\epsilon_\alpha^\star(k)\n
&=-ig^2Q_{EM}\bar{u}^w(p_3)\gamma_\nu  u^s(p_1)\frac{\lambda^a_{il}\lambda^a_{mj}}{q^2}\bar{u}^r(p_4)\n
&\quad\times\left[  \frac{\gamma^\nu\slashed{k}\gamma^\alpha-\gamma^\nu p_2^\alpha}{2p_2\m k }   +    \frac{\gamma^\alpha\slashed{k}\gamma^\nu+2\gamma^\alpha p_4^\nu}{2p_4\m k}     \right]u^t(p_2)\epsilon_\alpha^\star(k).
\end{align}
\end{widetext}
This amplitude needs to be squared in the next step, $(i\mathcal{M}_a+i\mathcal{M}_b)\m(i\mathcal{M}_a+i\mathcal{M}_b)^\star$,
 and then summed over final states and averaged over initial states. We define the resulting summed and averaged squared matrix element as $\overline{\lv\mathcal{M}\rv^2}$.
The sum over final photon polarizations reduces to \cite{Peskin:1995ev}
\begin{align}
\sum\limits_\epsilon \epsilon_\alpha^\star(k)\epsilon_\beta(k) \rightarrow -g^{\alpha\beta}.
\end{align}
The color matrices are (see Ref.~\cite{Peskin:1995ev}, Eq.~(17.63)) 
\begin{align}
\frac{1}{N_c^2}\sum\limits_{\text{colors}} \lambda^a\lambda^a\lambda^b\lambda^b = \frac{2}{9}.
\end{align}
The average over initial quark spins and sum over final spins gives a factor $1/4$, and, by using 
\begin{equation}
\sum\limits_{\text{spin }t}u^t(p)\bar{u}^t(p)=\slashed{p},
\label{eq:general_spinor_sum}
\end{equation}
we can transform the matrix element into traces,
\begin{widetext}
\begin{align}
\overline{\lv\mathcal{M}\rv^2}_{\text{rad.}} &= \frac{1}{4}\frac{2}{9}\frac{Q_{EM}^2g^4}{q^4}\text{Tr}\left\lbrace \slashed{p}_4  \left[
\frac{-\gamma^\nu \slashed{k} \gamma_\beta+2\gamma^\nu p_{2,\beta}}{2p_2\m  k}
+\frac{-\gamma_\beta\slashed{k}\gamma^\nu -2\gamma_\beta p_4^\nu}{2p_4\m k}\right]\right. \n
&\quad\times \slashed{p}_2
\left.
\left[ 
\frac{\gamma^\beta\slashed{k}\gamma^\mu-2\gamma^\mu p_2^\beta}{2p_2\m k}    
+\frac{\gamma^\mu \slashed{k}\gamma^\beta+2\gamma^\beta p_4^\mu}{2p_4\m k}   
\right]  \right\rbrace.
\label{eq:finalTrace}
\end{align}
The gluon momentum squared is $q^2=(p_4-p_2+k)^2$ and the gluon propagator reads,
\begin{align}
\frac{1}{q^4}=\frac{1}{(2p_4\m k -2k\m p_2-2p_4\m p_2)^2},
\end{align}
and after screening,
\begin{align}
\frac{1}{q^4}\rightarrow \frac{1}{\kl 2p_4\m k -2k\m p_2-2p_4\m p_2-m_{D,g}^2\kr^2}.
\end{align}
 The trace in Eq.~\eqref{eq:finalTrace} can be done using the {\sc mathematica} package {\sc FeynCalc 8.2.0} \cite{MERTIG1991345}, with the result (where we defined the scalar product of four-vectors $(ij)\equiv p_i\m p_j$),
\begin{align}
A&\equiv  2 (25)+m_{D,q}^2\n
B&\equiv  2 (45)+m_{D,q}^2\n
C&\equiv 4 (45)+m_{D,q}^2\n
D&\equiv (35) B^2-2 (34) A (2 (25)-B-m_{D,q}^2) \n
E&\equiv  (23) A ((25) C+(45) (-B-m_{D,q}^2))+(24) A (2 (34) A+(35) (A+B))+(25) D \n
F&\equiv  (24) A (A+B)+(25) B^2\n
G&\equiv  (23) A ((24) B+(45) A)+(34) F\n
H&\equiv  -2 (23) B+(34) (-B-m_{D,q}^2)+(35) m_{D,q}^2\n
J&\equiv  (45) H +(24) (35) B+(25) ((34) C+2 (35) (45))\n
  \overline{\lv\mathcal{M}\rv^2}_{\text{rad.}} &=\frac{1}{4}\frac{2}{9}Q_{EM}^2g^4 128\frac{A ((12) J-2 (13) (24) (45) A)+(14) E +(15) G}{A^2 B^2 (2 (24)+2 (25)-2 (45)+m_{D,g}^2)^2}
\end{align}
We have checked that the Ward identity is fulfilled.
\end{widetext}
\subsection{Symmetry-factor}
The self energy in Fig.~\ref{pic:photonSelfEnergy} with the given cut generates the $qq\rightarrow qq \gamma$ -process. We discuss its multiplicity factor here. The photon legs of the self energy can be crossed, which is why the self energy carries a factor of two. The four gluon vertices are completely identical. Every gluon can be reversed. This contributes a factor of four. The loop can be opened by tic-ing the upper or lower quark line, which introduces a factor of two. In total the symmetry factor is $16$.
\section{Verification of the Bremsstrahlung Process and Kinematics}
\label{sec:verification}
\begin{figure}
	\centering
	\includegraphics[width=0.95\columnwidth]{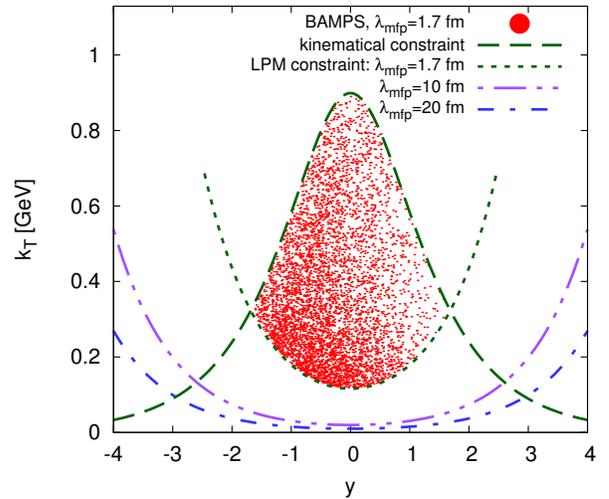}
	\caption{The exact photon bremsstrahlung  matrix element is used to sample photons. Their momentum is given in $k_\perp,q_\perp,y,\phi$-space; here we show several realisations (red dots) as an example. The green dashed curves represent the limits. The purple and blue dash-dotted lines show the limit from the LPM constraint for larger mean-free paths. The asymmetry in $y$ is forced by using only one fixed quark as the radiating one.  \label{pic:BAMPS_ktY}}
\end{figure}

\begin{figure}
\centering
\includegraphics[width=0.95\columnwidth]{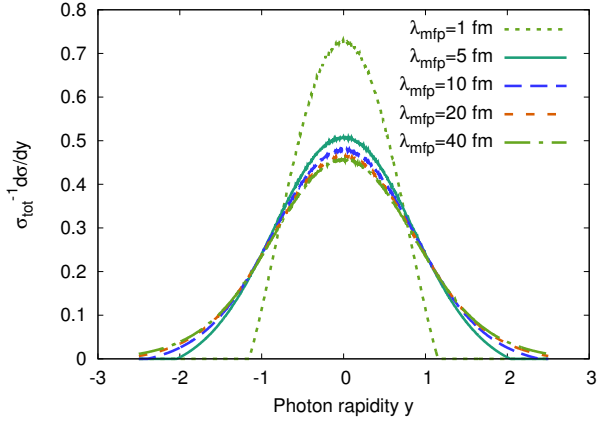}
\caption{The differential cross section in the rapidity of the radiated photon for various mean-free paths.
	\label{pic:BAMPS_yDiffCS}}
\end{figure}

\begin{figure}
\centering
\includegraphics[width=0.95\columnwidth]{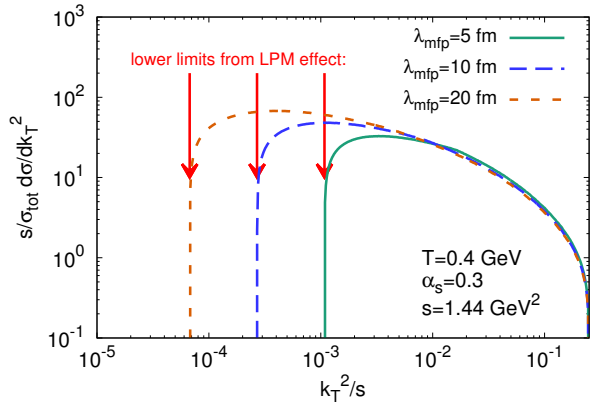}
\caption{The differential cross section in the transverse momentum of the radiated photon for various mean-free paths.
\label{pic:BAMPS_ktDiffCS}}
\end{figure}

To cross-check the inelastic photon production and verify the kinematic integration limits and the limit stemming from the LPM constraint, we show in Fig.~\ref{pic:BAMPS_ktY} a typical set of sampled photon momenta according to the full bremsstrahlung matrix element. Each of the red dots represents one sampled photon for a fixed (but arbitrary) configuration of incoming quark momenta. For illustrational purposes (to see the intrinsic asymmetry in $y$) we fix the quark line where the photon is emitted, and discard the radiation from the other quark line. 
However, in any real simulation of BAMPS the incoming quarks are randomly taken to be either quark one or two - thus the momentum spectrum will be symmetric in $y$. 
Omitting the integration over $y$ in Eq.~\eqref{eq:sigma23Tot}, we compute numerically the differential cross section $\dd \sigma_{23}/\dd y$ normalized by the total cross section $\sigma_{\text{tot}}$ in Fig.~\ref{pic:BAMPS_yDiffCS}. Here the symmetry in $y$ can clearly be seen.
Omitting the integration over $k_\perp^2$ in Eq.~\eqref{eq:sigma23Tot}, we compute also the differential cross section with respect to $k_\perp^2$, as shown in Fig.~\ref{pic:BAMPS_ktDiffCS}. Both figures are done for an arbitrary momentum setup of the incoming quarks, namely $p_1=(2T,0,2T,0), p_2=(2T,0,0,-2T)$ and $T=0.4~\mathrm{GeV}$. It is clearly visible, that the mean-free path changes the kinematics of the outgoing photon momenta, a larger mean-free path allows more collinear radiation.
%%%%%%%%%%%%%%%%%%%%%%%%%%%%%%%%%%%%%%%%%%%%%%%%%%%%%%%%%%%%%
%%%%%%%%%%%%%%%%%%%%%%%%%%%%%%%%%%%%%%%%%%%%%%%%%%%%%%%%%%%%%
% B I B L I O G R A P H Y
%%%%%%%%%%%%%%%%%%%%%%%%%%%%%%%%%%%%%%%%%%%%%%%%%%%%%%%%%%%%%
%%%%%%%%%%%%%%%%%%%%%%%%%%%%%%%%%%%%%%%%%%%%%%%%%%%%%%%%%%%%%
%\bibliographystyle{ieeetr} 
\bibliographystyle{apsrev4-1}
\bibliography{library_manuell.bib}

\end{document}